\documentclass[10pt,journal,compsoc]{IEEEtran}
\ifCLASSOPTIONcompsoc
  \usepackage[nocompress]{cite}
\else
  \usepackage{cite}
\fi
\ifCLASSINFOpdf
\else
\fi
\usepackage{amsmath,amssymb,amsfonts}
\newtheorem{thm}{Theorem}

\newtheorem{defn}[thm]{Definition} % 这句定义使得defn 环境和thm 共享编号

\usepackage{array}
\usepackage[ruled,vlined]{algorithm2e}
\usepackage{algorithmic}
\usepackage{graphicx}
\usepackage{subfigure} 
\usepackage{textcomp}
\usepackage{color} 
\usepackage{soul}
\usepackage{multirow}
\usepackage{amsfonts}
\usepackage{xcolor}
\usepackage{enumerate}
\definecolor{blue}{RGB}{0,0,0}
\definecolor{red}{RGB}{223,58,45}
\definecolor{green}{RGB}{34,139,34}

\begin{document}
\title{Traffic Load-Aware Resource Management Strategy for Underwater Wireless Sensor Networks}

\author{Tong Zhang,~\IEEEmembership{Member,~IEEE,}
	Yu Gou,~\IEEEmembership{Member,~IEEE,}
        Jun Liu,~\IEEEmembership{Member,~IEEE,}
        %Shanshan Song,
        %Jifeng Zhu,
        %Tingting~Yang,~\IEEEmembership{Member,~IEEE},
        and~Jun-Hong~Cui% <-this % stops a space
\IEEEcompsocitemizethanks{
\IEEEcompsocthanksitem Tong Zhang and Yu Gou are with Beihang Ningbo Innovation Research Institute, Beihang University, Ningbo, China, 315800, and also with the School of Electronic and Information Engineering, Beihang University, Beijing, China, 100191. E-mail: \{tong\_zhang, gouyu\}@buaa.edu.cn.
\IEEEcompsocthanksitem Jun Liu is with the School of Electronic and Information Engineering, Beihang University, Beijing, China, 100191. E-mail: liujun2019@buaa.edu.cn.
\IEEEcompsocthanksitem Jun-Hong Cui is with Shenzhen Institute for Advanced Study, UESTC, Shenzhen, China, and also with the College of Computer Science and Technology, Jilin University, Changchun, China.
}
\thanks{(Corresponding author: Jun Liu.)}
}


\markboth{Journal of \LaTeX\ Class Files,~Vol.~14, No.~8, August~2015}%
{Shell \MakeLowercase{\textit{et al.}}: Bare Demo of IEEEtran.cls for Computer Society Journals}



\IEEEtitleabstractindextext{%
\begin{abstract}
Underwater Wireless Sensor Networks (UWSNs) represent a promising technology that enables diverse underwater applications through acoustic communication. However, it encounters significant challenges including harsh communication environments, limited energy supply, and restricted signal transmission. This paper aims to provide efficient and reliable communication in underwater networks with limited energy and communication resources by optimizing the scheduling of communication links and adjusting transmission parameters (e.g., transmit power and transmission rate). The efficient and reliable communication multi-objective optimization problem (ERCMOP) is formulated as a decentralized partially observable Markov decision process (Dec-POMDP). A \textcolor{blue}{\underline{T}raffic Load-\underline{A}ware} \underline{R}esource \underline{M}anagement (TARM) strategy based on deep multi-agent reinforcement learning (MARL) is presented to address this problem. \textcolor{blue}{Specifically, a traffic load-aware mechanism that leverages the overhear information from neighboring nodes is designed to mitigate the disparity between partial observations and global states. Moreover, by incorporating a solution space optimization algorithm, the number of candidate solutions for the deep MARL-based decision-making model can be effectively reduced, thereby optimizing the computational complexity.} Simulation results demonstrate the adaptability of TARM in various scenarios with different transmission demands and collision probabilities, while also validating the effectiveness of the proposed approach in supporting efficient and reliable communication in underwater networks with limited resources.
\end{abstract}

}

\maketitle

\IEEEdisplaynontitleabstractindextext

% \fi
\IEEEpeerreviewmaketitle

\IEEEraisesectionheading{\section{Introduction}\label{sec:intro}}

\IEEEPARstart{C}{onsisting} of nodes with sensing and communication abilities, Underwater Wireless Sensor Networks (UWSNs) are recognized as one of the most promising technologies that facilitates various applications, including ocean environment monitoring, data collection, underwater resource exploration, and underwater ecological environment protection \cite{luo2021survey}\cite{gou2022deep}. Similar to terrestrial radio networks, UWSNs rely on wireless communication to establish network connectivity. However, the performance of UWSNs lags significantly behind terrestrial radio networks, which can be attributed to several reasons.

First, in the context of UWSNs, communication mainly occurs through acoustic channels because radio signals experience rapid attenuation in water \cite{gou2024achieving}. The underwater acoustic channel presents several challenges, including limited bandwidth, slow propagation speed (approximately 1500 m/s), and time-varying multi-path propagation \cite{morozs2020channel}. These characteristics collectively make underwater acoustic channel one of the most demanding and challenging mediums in use today \cite{stojanovic2009underwater}.

Second, the battery-powered nodes presents a major challenge in UWSNs. Due to the high costs and harsh operating environments, it is impractical to recharge or replace the batteries after deployment \cite{islam2022lifetime}. Consequently, the energy supply for underwater nodes is severely limited, acting as a bottleneck for UWSNs. Efficient energy utilization strategies are essential to extend the network lifetime, especially for long-term applications such as ocean observation networks. In addition, underwater nodes are typically equipped with half-duplex acoustic modems, indicating that a node can only receive one packet at a time and cannot transmit data while receiving \cite{su2020optimal}. This limitation further exacerbates network performance and channel utilization.

Third, to mitigate packet collisions in real-world scenarios, a round-robin transmission mechanism is often employed \cite{jiang2023medium}. This mechanism assigns each node specific time slots for collision-free communication. However, fixed duration of transmission slots does not effectively consider the unique characteristics of underwater environment or the specific requirements of underwater applications, leading to a significant reduction in channel utilization. Therefore, it is imperative to develop link scheduling algorithms tailored for underwater environments to enhance network performance and reliability while operating within the constraints posed by limited available resources.

Recent studies have highlighted the heterogeneity of UWSNs in terms of traffic load \cite{enhos2022coexistence}\cite{su2023hcar}. For many underwater applications like disaster monitoring and early warning networks, transmission demands from nodes are not evenly distributed in time and space, resulting in an extensive dynamic range of traffic load \cite{su2021traffic}. Allocating an equal number of transmission slots with identical durations to all nodes would result in inefficient utilization of communication resources. Nodes without data to transmit would unnecessarily consume valuable resources, while nodes with heavy data loads may experience significant delays due to insufficient transmission opportunities, decreasing overall network efficiency.

Proposed as practical solutions to address the mismatch between link scheduling and transmission demands in underwater networks, traffic load-aware transmission mechanisms are motivated by two primary considerations. On the one hand, the transmission demands for all nodes should be coordinated adequately to avoid collisions at the receiver. On the other hand, transmission parameters adapt to the dynamic network traffic load state to achieve higher channel utilization and energy efficiency. In heavy traffic load scenarios, scheduled nodes should minimize transmission delay while ensuring communication quality. Conversely, lower rates can conserve energy and prevent collisions in low-traffic and high-interference situations. Static allocation strategies that do not consider traffic load variations will reduce channel utilization and energy efficiency.

The current research has focused on centralized network architectures, wherein a central node gathers transmission demands from multiple transmitters and organizes their transmission slots through control signals. This centralized approach enables precise link scheduling based on load conditions. However, UWSNs are naturally sparsely distributed networks and require a distributed resource management strategy. The frequent exchange of control information leads to reduced channel utilization due to high propagation delays. Additionally, the challenges associated with centralized scheduling and communication overhead pose scalability issues as the number of network nodes increases \cite{zhang2021udarmf}. These limitations ultimately impact the effectiveness of centralized methods in underwater communication systems.

An alternative solution involves incorporating traffic load information into the header of each data packet \cite{hu2010qelar}. This allows the node to monitor transmissions within its communication range and acquire knowledge about the transmission demands of neighboring nodes. With this mutual perception capability, underwater nodes can collaboratively optimize network performance by determining their transmission behaviors in a distributed manner. Distributed solutions are often more suitable for underwater environments compared to centralized approaches. However, distributed solutions also face specific challenges. Firstly, including load information in packet headers introduces additional communication overhead. Secondly, underwater nodes can only perceive the load of neighboring nodes, which means they make decisions based on incomplete information. The mismatch between local observations and global states can affect the effectiveness of collaboration between nodes. Moreover, battery-powered underwater nodes must balance communication quality and energy consumption through environmental perception, which is crucial for maximizing long-term network performance while operating under limited energy resources.

This paper presents a Traffic Load-Aware Resource Management (TARM) strategy to achieve efficient and reliable communication in UWSNs. The TARM strategy aims to maximize the number of successful communications while minimizing failures caused by conflicts or interference. To address the efficient and reliable communication multi-objective optimization problem (ERCMOP) within the considered system, this paper formulates it as a decentralized partially observable Markov decision process (Dec-POMDP) \cite{nair2005networked} and proposes a deep multi-agent reinforcement learning (MARL) solution to solve it. Historical observations are leveraged to reduce decision uncertainty by bridging the gap between partial observations and global states.

To alleviate communication overhead, underwater nodes only attach necessary information to the data packets. Each node can gather load information about its one-hop neighbors by listening to network traffic and overhearing packets that were not originally meant for it. Following the guidelines outlined by TARM, nodes can transmit data at their own discretion without negatively impacting the network's overall performance. This flexibility in data transmission leads to significant improvements in channel utilization, throughput, and delivery ratio while also reducing the associated end-to-end delay and average energy consumption. The main contributions of this paper are as follows:
\begin{enumerate}[1)]
\item Enable efficient and reliable communication in UWSNs that operate in challenging acoustic channels and face constraints in terms of energy supplies and communication resources. To achieve this, the presented approach concentrates on two main components: scheduling the communication links between underwater nodes and adjusting the transmission parameters of individual nodes.
\item Formulate the multi-objective problem of efficient and reliable communication (ERCMOP) in UWSNs. The objective is to increase the number of successful communications and reduce transmission failures that result from collisions or interference simultaneously. Specifically, ERCMOP is addressed by TARM, which is a deep MARL-based traffic load-aware resource management strategy presented in this paper.
\item Design the traffic load-aware mechanism that leverages the overhear information obtained from neighboring nodes and the long propagation delay of the acoustic channel to mitigate the gap between partial observations and global states, thus enabling a more suitable link scheduling scheme and enhancing network performance. Furthermore, a solution space optimization algorithm is incorporated to optimize the computational complexity associated with the deep MARL-based strategy. This algorithm effectively reduces the number of candidate solutions, facilitating more efficient solution space exploration during the MARL training process.
\item Conduct extensive evaluations with respect to the real acoustic modem settings, specifically the transmission mode and transmit power. These evaluations aimed to assess the performance and adaptability of the TARM strategy in various scenarios with different transmission demands and collision probabilities. The simulation results confirmed the effectiveness of TARM in adapting to these diverse scenarios. The evaluations also focused on validating the critical design components of TARM. Based on the evaluation results, recommendations were provided for customizing TARM in real deployments.
\end{enumerate}

The remainder of the paper is organized as follows. In Section \ref{sec:rw}, we review the related work on underwater resource management strategies. We describe the system model and formulate the problem in Section \ref{sec:sys}. The TARM strategy is presented in Section \ref{sec:TARM_method}. Section \ref{sec:TARM_eva} gives the evaluation results. Finally, Section \ref{sec:con} concludes the paper.

\section{Related Work}\label{sec:rw}

This section provides a brief review of several studies related to underwater resource management strategies, focusing on link scheduling, power allocation, and rate adaptation. Additionally, recent research on traffic load-aware resource management strategies is also included.

In UWSNs, most nodes are battery-powered, which leads to a limited network lifetime and system performance. Many research efforts have focused on optimizing energy efficiency while maintaining performance \cite{hu2010qelar}. One promising approach to improve system performance while conserving energy is power control. The transmitter adjusts its power to ensure that the signal strength at intended receiver remains at a pre-specified level whenever possible or shuts down when channel conditions deteriorate beyond a certain point \cite{qarabaqi2011adaptive}\cite{signori2021geometry}. By choosing the appropriate transmit power, Chen \textit{et al.} simultaneously satisfy the signal strength requirements and minimize mutual interference among underwater nodes \cite{chen2022collision}. In \cite{zhou2008energy}, a cross-layer multi-path power control solution is presented to jointly reduce energy consumption, end-to-end delay, and packet error rate for time-critical applications. Additionally, Jin \textit{et al.} revealed that a multi-leader multi-follower Stackelberg game-based distributed power allocation solution helps achieve a better tradeoff between network throughput and system complexity \cite{jin2023joint}.

As the number of nodes that share the acoustic channel increases, substantial mutual interference among underwater nodes cannot be overcome by power control alone \cite{elbatt2004joint}. Joint link scheduling and power control have emerged as viable solutions in UWSNs for collision avoidance and performance optimization \cite{bai2015link}. In \cite{sun2021collision}, Sun \textit{et al.} avoid collisions by allocating different time slots for nodes in the same collision area while enabling spatial reuse for nodes in different collision areas. Their proposed strategy effectively reduces transmission delay and enhances energy efficiency across various network topologies. Leveraging long propagation delays to enable concurrent transmissions, Zhuo \textit{et al.} improve network throughput \cite{zhuo2019delay}. Zhang \textit{et al.} proposed a deep MARL-based joint link scheduling and power allocation scheme, which adapts the transmission behaviors of individual nodes to time-variant acoustic channel and dynamic network topology, seeking the optimal trade-off between network capacity and communication fairness \cite{zhang2024joint}.

Rate adaptation, which involves adjusting the modulation and coding scheme based on link quality, is recognized as an effective solution for improving network throughput, particularly in the presence of varying interference levels \cite{santagati2014medium}\cite{teixeira2015ieee}. As mentioned in \cite{wan2014adaptive} and \cite{ahmad2019effective}, higher transmission rates rely on a favorable communication environment characterized by stronger signal strength, while lower rates may compromise throughput and channel utilization but increase system robustness. Adjusting the transmission rate usually requires information about the channel state and the transmit power being used.

Network traffic fluctuates due to non-uniform transmission demands from nodes in both time and space dimensions, and the number of underwater nodes in the network also varies \cite{zhang2019load}. As revealed in \cite{chen2023time}, fluctuations in traffic load significantly affect network performance. Similarly, \cite{yang2023traffic} demonstrated through sea trials that considering variations in traffic load can greatly improve network performance. Recently, traffic load-aware resource management strategies have emerged as practical solutions for optimizing overall network performance in underwater acoustic networks.

Existing strategies involve nodes sharing their traffic load information with neighboring nodes through control packets \cite{zhuo2019delay}, or attaching metadata to the data packets they send \cite{hu2010qelar}. In \cite{su2021traffic}, Su \textit{et al.} proposed a distributed traffic load-aware link scheduling scheme for UWSNs, where the varying traffic load is measured by evaluating the proportion of time that the control channel is busy during the observation period. Control packets are exchanged to update the channel resource list of each node. In \cite{hsu2009st}, Hsu \textit{et al.} proposed a heuristic link scheduling algorithm (TOTA) that considers traffic load and routing information to save energy and improve throughput. The base station acts as a central controller, collecting information such as topology, traffic load on each link, propagation delay between node pairs, and interference relationships. With this information, the base station computes the optimal schedule. These strategies necessitate frequent information exchange between underwater nodes and the central controller, consuming limited bandwidth. Furthermore, they assume that all information is always available, which is not compliant with the unreliable and resource-constrained characteristics of acoustic channels.

In \cite{deng2018dco}, the network is divided into subnets based on traffic load, and an appropriate link scheduling scheme is adopted for each subnet. This method successfully reduces end-to-end delay and energy overheads. However, the interference among different subnets and within subnet nodes is not adequately addressed, leading to reduced throughput. Subsequently, in \cite{zhang2019load}, Zhang \textit{et al.} proposed a load-based time slot allocation scheme to mitigate intra-network interference and improve channel utilization under fluctuating traffic loads. However, both methods assume a fixed propagatioßn speed of acoustic wave and a static communication environment, which may compromise the effectiveness of these approaches in real deployments. Conversely, our presented traffic load-aware resource management strategy considers the dynamic nature of the acoustic channel while avoiding an over-reliance on neighboring load information, making it more suitable for UWSNs.

\section{System Model}\label{sec:sys}

\subsection{Network Model}\label{sec:system}

\begin{figure}[htbp]
\begin{center}
\includegraphics[width=3in]{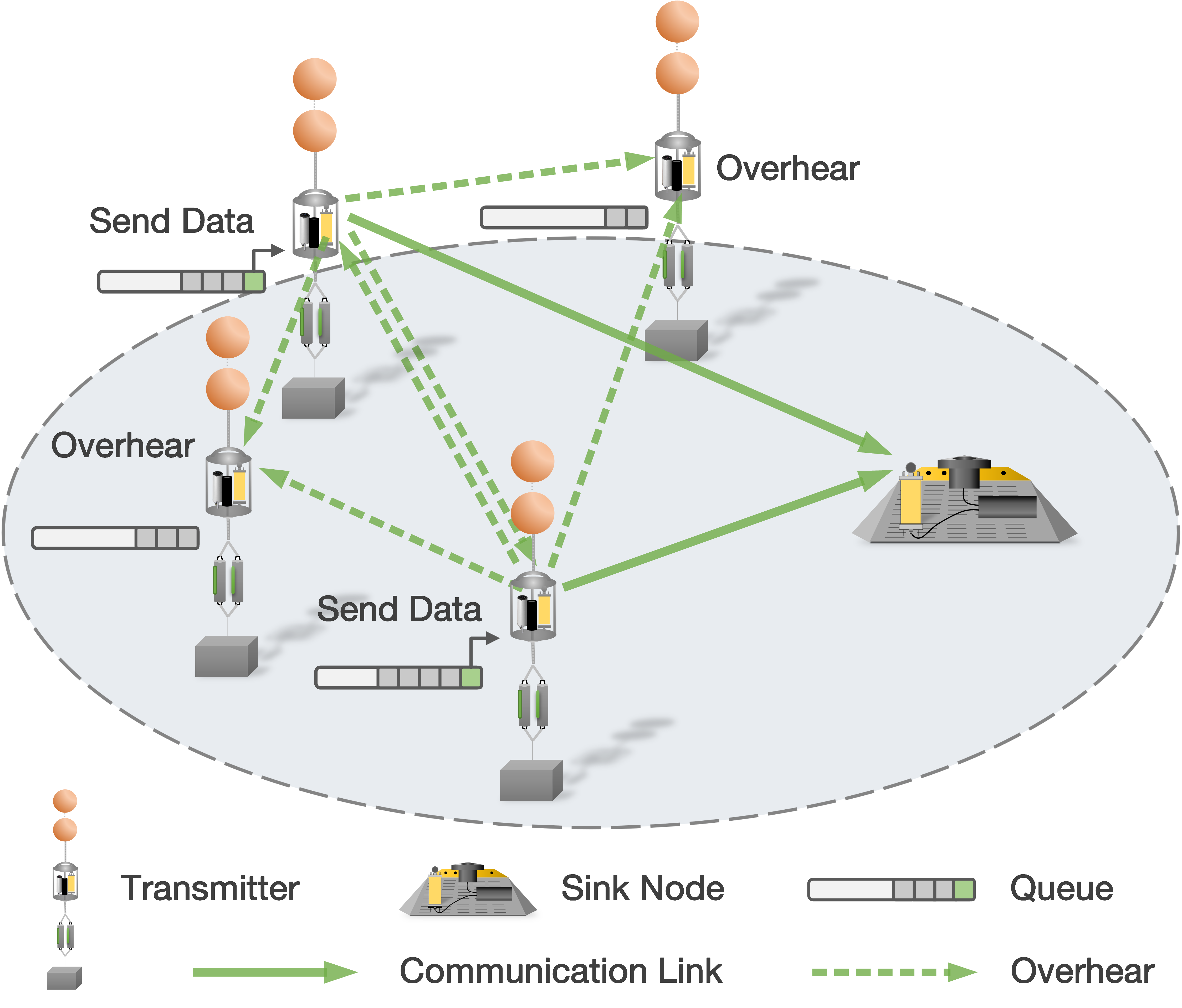}
\caption{The UWSNs considered in this paper.}
\label{fig:TARM_topology}
\end{center}
\end{figure}

This study focuses on the analysis of single-hop UWSNs consisting of $N$ transmitters that send data packets to a sink node through the shared acoustic channel, as shown in Fig. \ref{fig:TARM_topology}. The set of transmitters is denoted as $\mathcal{N}=\{1,2,...,N\}$, while the sink node is represented by $m$. All nodes in the underwater network are equipped with half-duplex and omni-directional acoustic modems. Furthermore, the clocks of all nodes are synchronized. At any given time $t$, \textcolor{blue}{the location of transmitter $n_i \in \mathcal{N}$ is denoted by $\boldsymbol{X}_{i}^{t}$, where $\boldsymbol{X}_{i}^{t}\!=\!(x_{i}^{t},y_{i}^{t},z_{i}^{t})$ is $n_i$'s 3D Cartesian coordinates at $t$ slot. The distance between transmitter $n_i$ and the sink node $m$ is given by $d_{i,m}^{t}=|\boldsymbol{X}_{i}^{t}-\boldsymbol{X}_{m}^{t}|$.} It is worth noting that underwater nodes move passively due to environmental factors such as ocean currents and tides. In this paper, we adopt the mobility model denoted as $\mathcal{M}_{k,c}$, which was presented in \cite{he2020trust}, to simulate the passive movement of the nodes, i.e., $\boldsymbol{X}_{i}^{t+1}=\mathcal{M}_{k,c}(\boldsymbol{X}_{i}^{t})$.

During each transmission in the network, the scheduled node sends a data packet with a size of $L_{d}$ bytes. Let $s_{i,m}(T_{ob})$ represent the number of packets sent from $n_i$ to $m$ during the observation period $T_{ob}$. Similarly, let $re_{i,m}(T_{ob})$ denote the number of packets successfully received at $m$. In most scenarios, the condition $re_{i,m}(T_{ob}) < s_{i,m}(T_{ob})$ holds, indicating that not all transmitted packets are successfully received. The equality $re_{i,m}(T_{ob}) = s_{i,m}(T_{ob})$ only holds when all transmissions from $n_i$ are successfully received at $m$. The amount of data successfully transmitted over $T_{ob}$ is calculated as (\ref{equ:TARM_i_byte}),
\begin{equation}
c_{i}(T_{ob})=\sum\nolimits_{j=1}^{re_{i}(T_{ob})}L_{d}^{j}
\label{equ:TARM_i_byte}
\end{equation}
where $L_{d}^{j}$ is the packet size of the $j$-th transmission. Considering the system with $N$ transmitters, the aggregate count of successful communications and the total number of bytes delivered successfully in $T_{ob}$ can be represented as (\ref{equ:TARM_N_success}) and (\ref{equ:TARM_N_byte}), respectively.
\begin{equation}
re_{\mathcal{N}}(T_{ob})=\sum\nolimits_{i=1}^{N}re_{i}(T_{ob})
\label{equ:TARM_N_success}
\end{equation}
\begin{equation}
c_{\mathcal{N}}(T_{ob})=\sum\nolimits_{i=1}^{N}c_{i}(T_{ob})
\label{equ:TARM_N_byte}
\end{equation}

\subsection{Acoustic Channel Model}\label{sec:channel}

The underwater acoustic channels are widely recognized as one of the most complicated communication mediums, which is featured by frequency-dependent attenuation, time-varying multi-path propagation, and long propagation delay due to the low acoustic propagation speed (about 1500 m/s) \cite{stojanovic2009underwater}. To ensure that the presented strategy is effective in real-world applications, this paper endeavors to construct a simulation environment that closely mirrors reality. By incorporating essential factors such as transmission loss, multi-path fading, ambient noise, and interference from simultaneous transmissions, we aim to narrow the disparity between simulation and reality. The signal strength at the receiver is measured by the Signal-to-Interference-plus-Noise Ratio (SINR), which is defined as the power of the intended signal divided by the sum of the interference power (from all the other interfering signals) and the power of ambient noise, denoted as $\gamma_{i,m}$ in this paper and calculated as (\ref{equ:TARM_sinr}) \cite{stojanovic2007relationship}: 
\begin{equation}
\gamma_{i,m}=\frac{\eta_{0}p_{i}h_{i,m}}{\eta_{0}\sum_{k=1, k \ne i}^{N}p_{k}h_{k,m}+N(f_c)\Delta f}
\label{equ:TARM_sinr}
\end{equation}
where $p_{i}$ and $p_{k}$ are the transmit power of $n_{i}$ and other concurrent transmitters $n_k \in \mathcal{N}^{-i}$, $\eta_{0}$ is the transducer efficiency when converting electrical to acoustic power. $h_{i,m}$ and $h_{k,m}$ are the channel gains of the communication links from $n_i$ and $n_{k \in \mathcal{N}^{-i}}$ to $m$, respectively. The channel gain is calculated as (\ref{equ:TARM_gain}),
\begin{equation}
h_{i,m}=H_{i,m}\rho^{2}
\label{equ:TARM_gain}
\end{equation}
where $\rho$ is the fading coefficient, which is a function of time and can be approximated by using a unit-mean Rayleigh distributed random variable with a cumulative distribution function expressed as $P[\rho\le x]\!=\!1\!-\!\exp(\frac{\pi x^{2}}{4})$ \cite{guan2020stochastic}. $H_{i,m}$ is described as the transmission loss experienced by a narrow-band-acoustic signal over a given spectrum, and can be described by the Urick propagation  as (\ref{equ:TARM_gain2}) \cite{urick1975principles},
\begin{equation}
H_{i,m}=(1000\times d_{i,m})^{-2}\cdot10^{-\frac{\alpha(f_c)d_{i,m}+A}{10}}
\label{equ:TARM_gain2}
\end{equation}
where $d_{i,m}$ in km is the distance between the $n_{i}$ and the sink node; $A \in [0,5]$ is the transmission anomaly in decibels, which accounts for the loss of acoustic intensity due to multi-path propagation, refraction, diffraction, and scattering of acoustic signals \cite{pompili2009cdma}; $f_c$ in kHz is the central carrier frequency; $\alpha(f_c)$ is the absorption coefficient, which can be calculated as \cite{stojanovic2007relationship}. $N(f_c)$ is the total power spectral density (PSD) in dB re $\mu$Pa per Hz of the ambient noise, which is composed of turbulent noise, transportation noise, thermal noise, and wave noise, as expressed in \cite{stojanovic2007relationship}. $\Delta f$ is a narrow band around frequency $f_c$.

The successful reception of the transmission is determined by the \textit{Physical Model} \cite{gupta2000capacity}, commonly known as the SINR model. This model is based on the realistic implementation of acoustic modems, where interference is considered as noise. According to the physical model, the transmitted signal can be decoded with an acceptable bit error rate (BER) if the SINR at the intended receiver exceeds a specified threshold $\gamma^{th}$; otherwise, the current transmission fails \cite{shi2012bridging}.

\subsection{Interference Model}\label{sec:interference_model}

The primary objective of this paper is to enhance network performance by effectively scheduling transmissions among underwater nodes, thereby minimizing collisions at the receiver. In order to provide a comprehensive understanding of the interference relationships among these nodes, the paper presents an interference model for the considered systems. This model serves to illustrate how different nodes may interfere with each other during transmission.

In the scenario where $n_i$ sends data packets to $m$, the overall delivery delay primarily consists of two components: the transmission delay $\delta^{tx}$, and the propagation delay $\delta^{prop}$. $\delta^{tx}$ is determined by the packet length $L_{d}$ and the transmission rate $r$, i.e., $\delta^{tx}=L_{d}/r$. On the other hand, the propagation delay $\delta^{prop}$ is influenced by the distance between nodes and the propagation speed of the acoustic signal. Considering the multi-path spread and the varying propagation speed, $\delta^{prop}$ is given by the BELLHOP \cite{porter2011bellhop} ray tracing algorithm.

Suppose that $n_i$ sends to the central node $m$ at time $t_i$, the arrival time $t_i^{arrive}$ of the data packet at $m$ can be represented as (\ref{equ:TARM_i_arrive}),
\begin{equation}
t_{i}^{arrive}=t_{i}+\delta_{i}^{prop}
\label{equ:TARM_i_arrive}
\end{equation}
and the completion time $t_{i}^{complete}$ at which $m$ finishes receiving the data packet from $n_i$ is expressed by (\ref{equ:TARM_i_complete}).
\begin{equation}
t_{i}^{complete}=t_{i}^{arrive}+\delta_{i}^{tx}=t_{i}+\delta_{i}^{prop}+\delta_{i}^{tx}
\label{equ:TARM_i_complete}
\end{equation}

Let $n_k \in \mathcal{N}$ represent another node with a data transmission requirement, where $i \neq k$. Similarly, we denote $t_{k}^{arrive}$ and $t_{k}^{complete}$ as the time when the data packet sent by $n_k$ arrives at $m$ and when $m$ completes receiving the data packet, respectively, and $t_k$ denotes the time when $n_k$ sends data to $m$, as shown in (\ref{equ:TARM_k_arrive}) and (\ref{equ:TARM_k_complete}).
\begin{equation}
t_{k}^{arrive}=t_{k}+\delta_{k}^{prop}
\label{equ:TARM_k_arrive}
\end{equation}
\begin{equation}
t_{k}^{complete}=t_{k}^{arrive}+\delta_{k}^{tx}=t_{k}+\delta_{k}^{prop}+\delta_{k}^{tx}
\label{equ:TARM_k_complete}
\end{equation}

In order to ensure conflict-free communication between nodes, the following conditions must be met: 1) If the data packet from $n_i$ arrives at $m$ before the data packet from $n_k$, then the arrival time of the data packet from $n_k$ must be after the completion time of the data packet from $n_i$; 2) If the data packet from $n_k$ arrives at $m$ first, then the arrival time of the data packet from $n_i$ must be after the completion time of the data packet from $n_j$. These constraints can be expressed mathematically as (\ref{equ:TARM_math_cons}).
\begin{equation}
t_{i}^{complete}<t_{j}^{arrive}\parallel t_{i}^{arrive}>t_{j}^{complete}
\label{equ:TARM_math_cons}
\end{equation}
The above constraints ensure that the data packets sent by different nodes arrive and complete their reception at $m$ sequentially and non-conflictingly.

\subsection{Energy Model}\label{sec:energy_model}

The initial battery capacity of the underwater nodes is denoted as $E^{0}$ and measured in joule (J). The total energy consumption of $n_{i}$ is represented by $E_{i}=\sum_{j=1}^{s_{i,m}(T_{ob})}e_{i}^{j}$, where $e_{i}^{j}$ signifies the energy consumption for the the $j$-th transmission by node $n_{i}$, calculated as $e_{i}^{j}=\int_{t=t_{i}^{j}}^{t_{i}^{j}+\delta_{i}^{tx,j}}p_{i}(t)dt$. This energy consumption is mainly affected by the transmit power $p_{i}$, data payload length $L_{d}$, and transmission rate $r$. As most underwater nodes are battery-powered, replacing or recharging the battery after deployment is challenging. The availability of energy supplies has a significant impact on network performance. Therefore, meeting the communication threshold while minimizing energy consumption is a desirable goal.

\subsection{Problem Formulation}\label{sec:pf}

This paper aims to enhance the throughput and reliability of the considered underwater system by increasing the number of successful communications in the network. One of the methods involves decreasing the transmission delay of individual packets by boosting the transmission rate of the underwater nodes \cite{wan2014adaptive}. However, a higher transmission rate typically requires increased transmit power to maintain network reliability and ensure the bit error rate (BER) remains below a certain threshold \cite{fan2021adaptive}. Due to the limited energy supplies of the battery-powered underwater nodes, it becomes necessary to dynamically adjust the transmission parameters to strike a balance between energy consumption and transmission delay. Moreover, increasing the number of successful transmissions at the receiver will further optimize the overall network throughput.

The optimization objectives mentioned above can be achieved through a combination of joint link scheduling and adjustment of transmission parameters. This paper considers practical acoustic modem designs, with transmission rate and transmit power of the underwater nodes being the key transmission parameters \cite{zhu2015toward}. Each node initially adjusts its transmission parameters to balance energy consumption and transmission delay, taking the communication environment into account. Afterward, a suitable number of underwater nodes are scheduled for transmission, considering the fluctuations in traffic load and their impact on available bandwidth and queueing delay \cite{chen2024time}. The optimization objectives of ERCMOP can be further categorized into three sub-objectives: maximizing network throughput, minimizing transmission delay, and minimizing energy consumption for packet transmission. Furthermore, the combination of modulation scheme and code rate, known as the transmission mode, is used to represent the node transmission rate in the subsequent sections \cite{wan2013field}.

\textit{1) Maximize network throughput.} Each node $n_{i} \in \mathcal{N}$ with a non-empty transmission queue aims to maximize network throughput by increasing the number of successful communications $re_{\mathcal{N}}(T_{ob})$ during the observation period $T_{ob}$, as indicated in (\ref{equ:TARM_N_success}). The first objective function can be formulated as follows:
\begin{equation}
f_{1}=re_{\mathcal{N}}(T_{ob})=\sum_{i=1}^{N}re_{i,m}(T_{ob})
\label{eq:TARM_obj1}
\end{equation}
In the underwater system being considered, successful transmissions must satisfy two constraints \cite{pompili2009cdma}. Firstly, the transmitted signal strength $\gamma_{i,m}$ must exceed the required communication threshold $\gamma^{0}$ in order for sink $m$ to correctly decode the signal. Secondly, there should be no conflict between any two communications at node $m$, satisfying the constraint depicted in (\ref{equ:TARM_math_cons}). An appropriate number of communication links are scheduled by being aware of the network traffic load. The nodes then achieve successful communications by jointly optimizing the transmission rate and transmit power at each allocated slot.

\textit{2) Minimize transmission delay.} Optimizing transmission delay is crucial for reducing the delivery delay of each data packet and for improving channel utilization. In this paper, channel utilization $U\in[0,1]$ is defined as the proportion of time during the observation period $T_{ob}$ that the shared acoustic channel is occupied for successful transmissions. The transmission delay of a single packet $\delta_{i,m}$ is determined by the packet size $L_{d}$ and the adopted transmission rate $r_{M}$. Specifically, $\delta_{i,m}$ is given by $\delta_{i,m}=L_{d}/r_{M}$, where $M \in \mathcal{M}$ represents the adopted transmission mode and $\mathcal{M}$ is the set of all available modes associated with the acoustic modem. By tuning the acoustic modem to a high-rate mode, $\delta_{i,m}$ can be reduced. System performance can be optimized by minimizing the transmission delay of each transmission in the network during $T_{ob}$. Therefore, the second objective function can be defined as follows:
\begin{equation}
f_{2}=\frac{\sum_{i=1}^{N}\sum_{j=1}^{s_{i,m}(T_{ob})}\delta_{i,m}^{j}}{\sum_{i=1}^{N}s_{i,m}(T_{ob})}
\label{eq:TARM_obj2}
\end{equation}
where $\delta_{i,m}^{j}$ is the duration for the $j$-th transmission of $n_i$. Higher transmission rates can reduce transmission delay and improve channel utilization and system performance during high traffic load.

\textit{3) Minimize energy consumption.} In order to achieve the objectives mentioned above, underwater nodes may operate in high-rate transmission modes with high transmit power to maximize network throughput while ensuring reliability. However, this increased performance comes at the cost of higher energy consumption for the battery-powered transmitters. To address this issue and extend network lifetime, it is important to reduce the average energy consumption across all transmitters. Therefore, the third objective function can be formulated as (\ref{eq:TARM_obj3}).
\begin{equation}
f_{3}=\frac{\sum_{i=1}^{N}\sum_{j=1}^{s_{i,m}(T_{ob})}e_{i,m}^{j}}{\sum_{i=1}^{N}s_{i,m}(T_{ob})}
\label{eq:TARM_obj3}
\end{equation}
where $e_{i,m}^{j}$ represents the energy consumption for the $j$-th transmission of $n_i$. The communication links need to be carefully scheduled, and the transmission parameters should be dynamically adjusted to avoid energy wastage from transmission collisions.

The ERCMOP problem involves three conflicting optimization objectives. Objective I aims to maximize the number of successful communications in the network without considering energy consumption and transmission delay. Objective II minimizes the average transmission delay across all data packets, disregarding energy consumption. Objective III seeks to reduce energy consumption, potentially compromising network throughput and increasing node delivery delay. Considering these three optimization objectives, the ERCMOP problem can be formulated as:
\begin{subequations}
\begin{align}
%\max\quad& \sum_{j=m}^{n}C_{j},\\
\min_{X}\quad& \{-f_{1},f_{2},f_{3}\}\\
\text{s.t.}\quad
&\text{C1: }m_{recv}^{t}\in\{0,1\}\label{equ:TARM_obj_hd}\\
&\text{C2: }\gamma_{i,m} \ge \gamma^{0}_{M}, i \in \mathcal{N} \label{equ:TARM_obj_th}\\
&\text{C3: }p_{i,m} \in [p_{min},p_{max}]\label{equ:TARM_obj_modem}
\end{align}
\label{equ:TARM_objmin}
\end{subequations}
\textcolor{blue}{where $X=[\mathbb{L},\mathbb{M},\mathbb{P}]$ represents the solution to ERCMOP. The matrix $\mathbb{L}^{N\times{T_{ob}}}=\{l_{i,m}^{t}|\forall i\in\mathcal{N}, \forall t\in{T_{ob}}\}$ denotes the scheduling states of all communication links, where $l_{i,m}^{t}$ indicates whether the communication link from $n_{i}$ to sink node $m$ is scheduled at the $t$-th observation slot. The matrix $\mathbb{M}^{N\times{T_{ob}}}=\{M_{i}^{t}|\forall i\in\mathcal{N}, \forall t\in{T_{ob}}\}$ represents the transmission modes of all transmitters, where $M_{i}^{t}$ denotes $n_{i}$'s transmission mode at the $t$-th observation slot. Similarly, $\mathbb{P}^{N\times{T_{ob}}}=\{p_{i}^{t}|\forall i\in\mathcal{N}, \forall t\in{T_{ob}}\}$ represents the transmit power of all transmitters, while $p_{i}^{t}$ denotes $n_{i}$'s transmit power.} Additionally, $m_{recv}^{t}$ denotes the number of communications received at node $m$ at any time $t \in T_{ob}$: $m_{recv}^{t}=1$ indicates that node $m$ is receiving, while $m_{recv}^{t}=0$ means that the receiver is idle at that time. When $m_{recv}^{t}>1$, it implies that conflicts are occurring at the receiving node, violating constraint (\ref{equ:TARM_math_cons}). Therefore, constraint (\ref{equ:TARM_obj_hd}) is used to constrain the rx-rx conflicts at the receiver. Constraint (\ref{equ:TARM_obj_th}) is employed to ensure communication quality. $\gamma^{0}_{M}$ is the communication threshold for transmission mode $M$. Given the communication distance and transmission mode, nodes should select an appropriate transmission power to meet the communication quality requirements. A high-speed transmission mode typically necessitates greater signal strength at the receiver to fulfill the decoding process. Constrained by the performance of the underwater acoustic modem, $p_{min}$ and $p_{max}$ respectively represent the minimum and maximum allowable transmit powers for the nodes. Constraint (\ref{equ:TARM_obj_modem}) restricts nodes from transmitting data with a power that exceeds their hardware limitations.

\section{Design of the TARM Scheme}\label{sec:TARM_method}

\subsection{Overview of TARM}\label{sec:TARM_description}

\begin{figure*}[htbp]
\centerline{\includegraphics[width=5.5in]{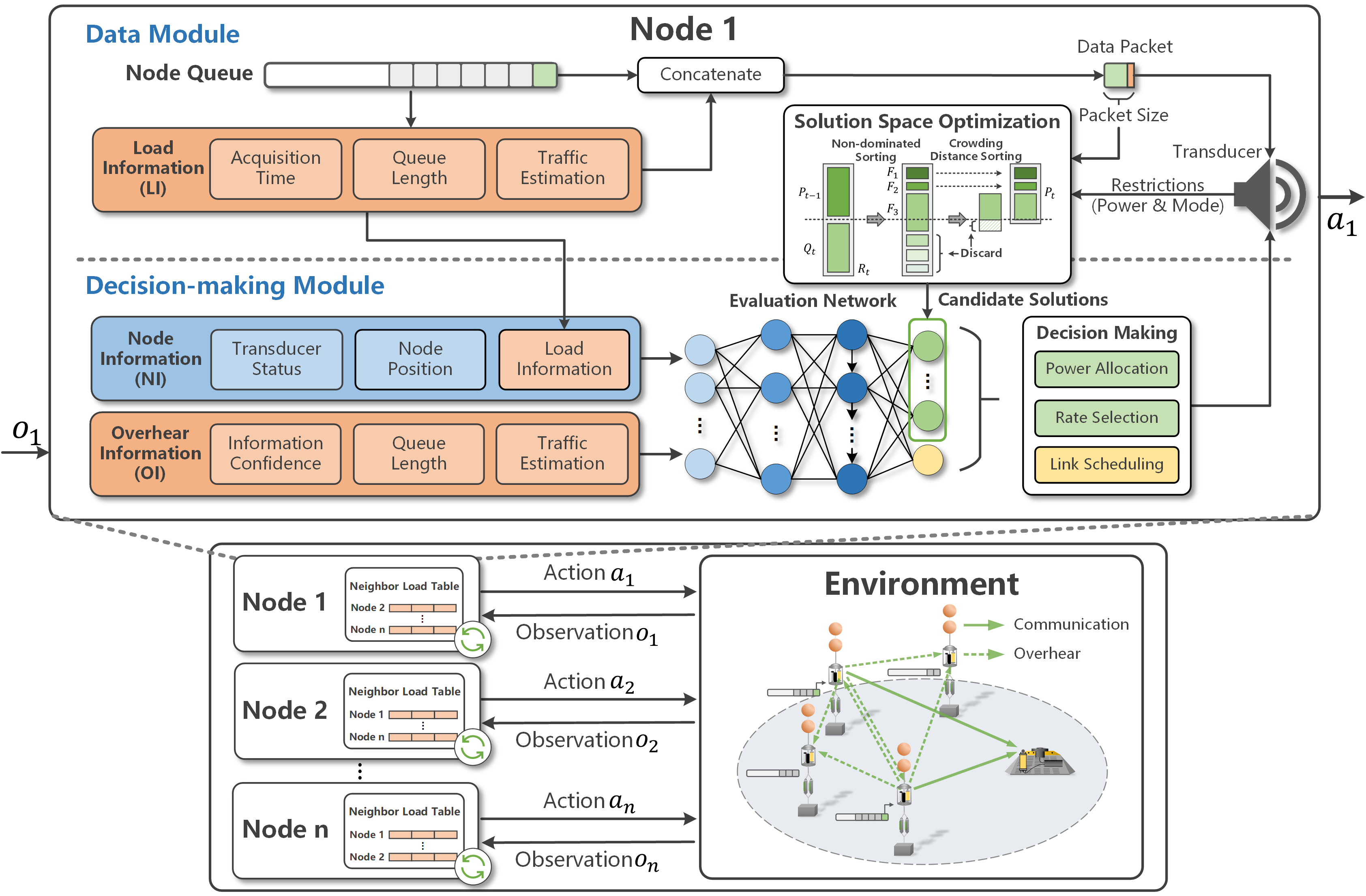}}
\caption{\textcolor{blue}{Overview of TARM.}}
\label{fig:TARM_arch}
\end{figure*}

To address the ERCMOP, a traffic load-aware resource management strategy, referred to as TARM, has been introduced. TARM facilitates efficient and reliable communication among underwater nodes by scheduling communication links and adjusting the transmission parameters of individual nodes. Within TARM, each underwater node utilizes partially observed traffic state information to generate an appropriate scheduling scheme. Specifically, by leveraging local traffic information and overheard data from neighboring nodes regarding their traffic loads, each underwater node independently determines its scheduling state and transmission parameters, which include transmission mode and transmit power. TARM employs deep multi-agent reinforcement learning (MARL) algorithms for its implementation and is trained using the centralized training with decentralized execution (CTDE) paradigm \cite{wang2020roma}. The workflow of TARM is illustrated in Fig. \ref{fig:TARM_arch}.

TARM's design tackles the following three challenges:
\begin{enumerate}
\item \textbf{Dynamic network traffic load state.} The individual underwater nodes experience heterogeneous and dynamically varying data loads due to variations in sampling frequencies specific to different applications. Consequently, transmission demands fluctuate across the underwater nodes. The mismatch between transmission opportunities and transmission demands renders traditional fixed link scheduling mechanisms inadequate for optimizing the overall performance of the network.
\item \textbf{Long propagation delay.} The underwater acoustic channel exhibits long propagation delays, leading to reduced channel utilization and the presence of outdated overhear traffic information. It is challenging to make the scheduling decisions that increases the channel utilization with outdated information.
\item \textbf{Huge solution space.} \textcolor{blue}{In ERCMOP problem, the scheduled nodes must balance their transmission mode and transmit power to simultaneously optimize all objectives, given a particular link scheduling decision. The size of the solution space, which corresponds to the action space for the deep MARL-based algorithm, increases with the number of available transmission modes and transmit power levels. This expansion presents a challenge for the deep MARL-based algorithm in terms of training efficiency and model complexity.}
\end{enumerate}

TARM involves two key processes to solve the ERCMOP problem. First, it utilizes the NSGA-II algorithm to optimize the solution space for ERCMOP, thereby reducing computational complexity during training and runtime. This process identifies the Pareto-optimal solutions, providing a set of efficient trade-offs between conflicting objectives. Then, TARM leverages a deep MARL algorithm to achieve joint link scheduling and adaptation of transmission parameters, enabling efficient and reliable communications in UWSNs. \textcolor{blue}{Specifically, a traffic load-aware mechanism is proposed to leverage the long propagation delays and enable concurrent transmission, while handling outdated overheard traffic information. This helps mitigate the gap between partial observations and global states.} The rest of this section is structured according to how TARM address the aforementioned challenges: $\S$\ref{sec:TARM_MOO} presents the solution space optimization (SSO) algorithm, $\S$\ref{sec:TARM_TA} introduces the traffic load-aware mechanism, and $\S$\ref{sec:TARM_DRLRM} presents the deep MARL-based traffic load-aware resource management strategy.

\subsection{Solution Space Optimization for ERCMOP}\label{sec:TARM_MOO}

In order to apply the deep MARL algorithm to address ERCMOP, it is imperative to define the state space, action space, and reward function for the algorithm. The action space encompasses various combinations of network transmission parameters, where each action within the space corresponds to a specific combination. Specifically, the action space in this paper consists of combinations of transmission mode and transmit power, i.e., $a=\langle M,p\rangle$. However, the large number of potential transmission combinations makes the action space high-dimensional, presenting a significant challenge for reinforcement learning algorithms to efficiently search for optimal solutions within such extensive spaces. Therefore, it is crucial to employ appropriate methods to reduce the size of the action space.

ERCMOP delineates a complex multi-objective optimization task, seeking an equilibrium across three pivotal goals: maximizing network throughput, minimizing packet transmission time, and minimizing energy consumption. Achieving maximum network throughput involves efficient scheduling of communication links. For the scheduled nodes, additional considerations are required regarding their transmission modes and transmit power to strike a balance between minimizing packet transmission time and minimizing energy consumption. Therefore, this section focuses on the above conflicting optimization objectives with decision variables being the transmission mode $\mathbb{M}$ and the transmit power $\mathbb{P}$, as defined in (\ref{equ:TARM_subobj1}). 
\begin{subequations}
\begin{align}
\min_{\mathbb{M},\mathbb{P}}\quad& \{\delta_{i,m}^{tx},e_{i,m}\}\\
\text{s.t.}\quad
&\text{C1: }\gamma_{i,m} \ge \gamma^{0}_{M}, i \in \mathcal{N} \label{equ:TARM_subobj1_th}\\
&\text{C2: }M \in \mathcal{M} \label{equ:TARM_subobj1_mode}\\
&\text{C3: }p_{i,m} \in [p_{min},p_{max}], i \in \mathcal{N}\label{equ:TARM_subobj1_power}
\end{align}
\label{equ:TARM_subobj1}
\end{subequations}
When dealing with an acoustic modem that offers multiple transmission modes $M \in \mathcal{M}$ and a continuous range of transmit power between $p_{min}$ and $p_{max}$, using all possible combinations as the action space for the reinforcement learning algorithm can lead to convergence difficulties or convergence to local optima. This paper adopts the Non-dominated Sorting Genetic Algorithm II (NSGA-II) \cite{deb2002fast} for optimizing the solution space in ERCMOP, which demonstrates good convergence and distribution characteristics in the obtained solution set, while also achieving high operational efficiency and acceptable computational complexity. As given in Algorithm \ref{alg:TARM_MOO}, NSGA-II combines genetic algorithms with Pareto dominance to optimize solutions through fundamental operations such as crossover, mutation, and selection \cite{pan2023joint}. Additionally, it incorporates a fast non-dominated sorting approach and a fast crowding-distance measurement method to enhance the efficiency, accuracy, and robustness of the algorithm \cite{yusoff2011overview}. Specifically, solutions within each non-dominated rank are sorted in descending order using the crowded-comparison operator $\prec_{n}$ defined in \cite{deb2002fast}. By employing NSGA-II, a set of non-dominated solutions (i.e., Pareto front) can be obtained, which helps reduce the size of the solution space significantly. 
\begin{algorithm}
\caption{SSO for ERCMOP.}\label{alg:TARM_MOO}
Initialize population size $N_{pop}$, number of generation $T_{gen}$, set of available transmission modes $\mathcal{M}$, range of transmit power $[p_{min}, p_{max}]$.\\
Initialize population $P_{0}=\{a_1, a_2, ..., a_{N_{pop}}\}$.\\
Initialize candidate solution set $\hat{\mathcal{U}} \gets \varnothing$.\\
Initialize non-dominated sorting $P_{0}=(\mathcal{F}_{1},\mathcal{F}_{2},...)$ and crowding-distance sorting for $P_{0}$.\\
\For{$j=1,...,T_{gen}$}{
Selection, crossover, and mutation to produce $Q_{j}$, which is the offspring combinations of $M$ and $p$ for the underwater node.\\
Combine parent population $P_{j-1}$ and offspring population $Q_{j}$, $R_{j}=P_{j-1}\cup Q_{j}$.\\
Calculate non-dominated fronts $\mathcal{F}=(\mathcal{F}_{1},\mathcal{F}_{2},...)$ of $R_{j}$, and assign crowding-distance in each $\mathcal{F}_{i} \in \mathcal{F}$.\\
\For{$\mathcal{F}_{i} \in \mathcal{F}$}{
Sort $\mathcal{F}_{i}$ in descending order using crowded-comparison operator $\prec_{n}$.
}
$P_{j}=\varnothing$.\\
\For{$\mathcal{F}_{i} \in \mathcal{F}$}{
\If{$|P_{j} \cup \mathcal{F}_{i}| \le N_{pop}$}{
$P_{j}=P_{j} \cup \mathcal{F}_{i}$.\\
\Else{
$P_{j}=P_{j} \cup \mathcal{F}_{i}[1:N_{pop}-|P_{j}|]$.\\
}
}
}
}
$\hat{\mathcal{U}}=\mathcal{F}_{1} \subseteq P_{T_{gen}}$.\\
\Return $\hat{\mathcal{U}}$.
\end{algorithm}

\begin{figure}[htbp]
\centerline{\includegraphics[width=3in]{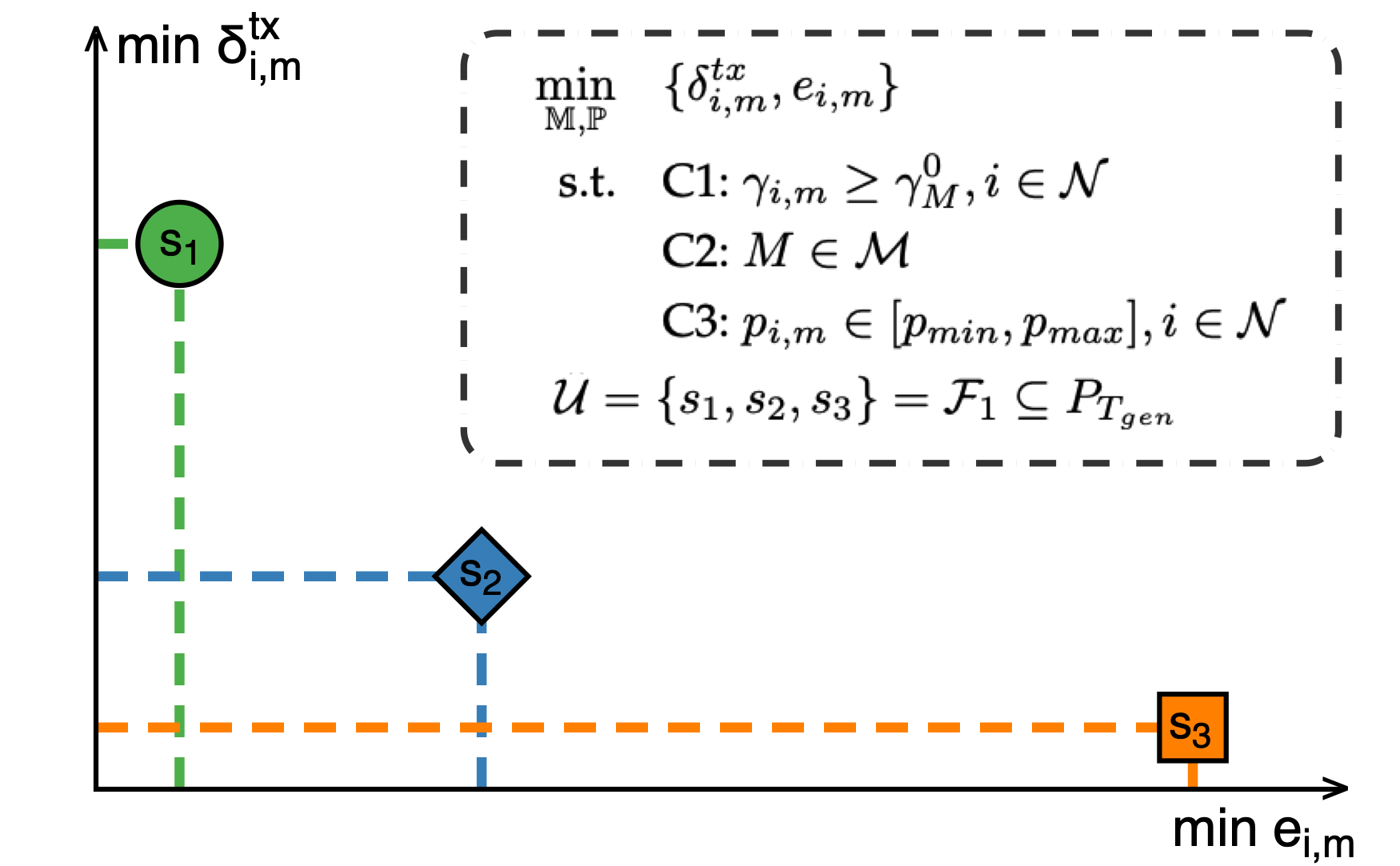}}
\caption{Sketch map of the candidate solution subset.}
\label{fig:TARM_MOOsketch}
\end{figure}

Using Fig.\ref{fig:TARM_MOOsketch} as an illustration, each point on the Pareto front represents a balanced solution between transmission time and energy consumption. The green solid circle corresponds to an action that minimizes energy consumption but has a longer transmission time. The orange solid square represents a case where the node prioritizes minimizing the transmission time, even if it results in higher energy consumption. The blue solid diamond, positioned between the green circle and the orange square, represents a solution that does not exclusively minimize either of the optimization objectives. Instead, it achieves a balance between the two objectives. All three solutions within the Pareto front are mutually non-dominated, indicating that no solution outperforms another concerning both optimization objectives simultaneously. 

After obtaining the Pareto front that contains non-dominated solutions, the network operator can select a suitable solution based on specific application requirements. In the following section,  we will use these non-dominated solutions from the Pareto front to create the action space for the reinforcement learning algorithm.

\subsection{Traffic Load-aware Mechanism}\label{sec:TARM_TA}

To better utilize the long propagation delay in underwater acoustic communication and improve channel utilization while avoiding collisions, this section suggests a traffic load-aware mechanism based on information confidence. Each node attaches its local load information to the data packets it sends. Since every node in the UWSN overhears the traffic, it can gather metadata about its neighbors from the incoming packets, even if the packets are not intended for it. The load information is inserted concurrently with the extraction of data packets from node queues. Below are the main definitions of the traffic load-aware mechanism.
\begin{defn}
(Local load information.) Local load information is presented by $LI_{i}=\langle t_{acquire}, L_{queue}, est\rangle$, $n_{i} \in \mathcal{N}$. $t_{acquire}$ represents the information acquisition time; $L_{queue}$ denotes the number of data packets waiting in queues; and $est$ refers to the estimated traffic trend obtained using a linear traffic estimator $g(\cdot)$, which estimates the average packet generation rate (pkt/s) of the most recent $\kappa$ packets in the queue.
\label{defn:TARM_LI}
\end{defn}
\begin{defn}
(Neighbor load information table.) For a node $n_{i} \in \mathcal{N}$, it maintains a neighbor load information table, denoted as $T_{NL}$, to store the latest load information of neighboring nodes. $row=|T_{NL}|$ represents the number of entries in the table. Each entry $NL_{j}=\langle t_{acquire}, L_{queue}, est\rangle$ in $T_{NL}$ corresponds to a piece of neighbor load information.
\label{defn:TARM_NL_table}
\end{defn}
\begin{defn}
(Information confidence.) \textcolor{blue}{The information confidence $CF$ measures the trustworthiness of the stored information in $T_{NL}$ when it is used as an input for the decision-making model. Let $\triangle\delta_{j}$ denote the duration between the information acquisition time and the current time when using $NL_{j}$ for decision-making. $\triangle\delta_{j}$ accounts for transmission delay, propagation delay, and queuing delay. $CF$ is calculated using the hyperbolic tangent function as (\ref{equ:TARM_CF}).
\begin{equation}
CF=\frac{e^{\frac{\triangle\delta}{a}}-e^{-\frac{\Delta\delta}{a}}}{e^{\frac{\Delta\delta}{a}}+e^{-\frac{\Delta\delta}{a}}}
\label{equ:TARM_CF}
\end{equation}
where $a$ is a time scale coefficient used to control the rate at which the information confidence function changes along the time axis. The resulting confidence value falls within the range of $(-1, 1)$.}
\label{defn:TARM_confidence}
\end{defn}

\textcolor{blue}{The TARM node uses the load information obtained from neighboring nodes to predict future changes in network traffic. This helps bridge the gap between local observations and global network states. When a node overhears a new data packet, it checks the source ID in the header against the existing entries in $T_{NL}$.  If load information for the corresponding neighboring node is already in $T_{NL}$, the previous record is updated with the most recent data. If there is no corresponding entry, a new row is added to accommodate the newly acquired load information.}

\textcolor{blue}{When the load information of neighboring nodes in $T_{NL}$ is used as input for the decision-making model, the TARM node assesses the confidence of each record. It then replaces the original entry's information acquisition time with the confidence value $CF$. Specifically, data entries exhibiting a $CF$ value closer to zero are given increased confidence in the decision-making process. Finally, the restructured overhear information, referred to as $OI$, is used as input for the model.} Algorithm \ref{alg:TARM_confidence} outlines the process of constructing the overhear information matrix.
\begin{algorithm}
\caption{Constructing the overhear information matrix.}\label{alg:TARM_confidence}
Initialize overhear information matrix $\mathbf{OL}\gets T_{NL}$.\\
\For{$t=1, ..., T$}{
\For{$j=1, ..., row$}{
Select $NL_{j}=\langle t_{j,acquire}, L_{j,queue}, est_{j}\rangle$ from $T_{NL}$.\\
Calculate the duration $\triangle\delta_{j}$  between the time of information acquisition and the current moment.\\
Calculate the load information confidence $CF_{j}$ according to (\ref{equ:TARM_reward_scale}).\\
$OI_{j}=\langle CF_{j}, L_{j,queue}, est_{j}\rangle$
}
}
\Return $\mathbf{OL}$.
\end{algorithm}

\subsection{Deep MARL-based Resource Management}\label{sec:TARM_DRLRM}

In this section, we design our TARM strategy with the deep MARL algorithms and the traffic load-aware mechanism. Specifically, we first associated the formulated ERCMOP with several key enabling components of RL algorithms (such as the action space, observation information, and reward function). Then, the training and execution procedures of TARM are presented. In section \ref{sec:TARM_MOO}, we construct the candidate solution set through the NSGA-II algorithm, which will be used for implementing the action space for our proposed deep MARL-based resource management strategy. It should be noted that the TARM strategy is not restrict to any specific multi-objective optimization method, as long as the method can rapidly and stably generate a set of non-dominated candidate solutions from the solution space.

This paper takes the underwater nodes as agents, the UWSNs as multi-agent systems, and everything beyond the agents as the environment. Due to the limited sensing and communication range, the underwater nodes are only able to make observations of local information but not global state. Therefore, the formulated ERCMOP is as a \textcolor{blue}{Dec-POMDP}.

The Dec-POMDP can be represented by the tuple $G=\langle S,A,T,Z,O,R,h\rangle$, where $S$ is the set of environment states, and the initial state is denoted by $s_{0}$. $A=\{A_{i}\}_{i\in\mathcal{N}}$ is the joint action space for all nodes, where $A_{i}$ represents the action space of node $n_i$. $T: S\times A\times S \in [0, 1]$ is the state transition function, with $T(s' | s, \mathbf{a})$ denoting the probability of transitioning from state $s$ to $s'$ when all nodes choose the joint action $\mathbf{a} \in A$. $O = \{O_i\}_{i\in\mathcal{N}}$ is the joint environmental observation made by all nodes, where $O_i$ is the observation of $n_i$. $R$ is the reward function, and $r=R(s, \mathbf{a})$ represents the immediate reward obtained by all nodes after executing the joint action $\mathbf{a} \in A$ in state $s$. Subsequently, we will provide a detailed introduction to the action set $A$, observation $O$, and reward function $R$ of TARM.

\textbf{1) Action space $A$.} Let the function $f_{SSO}(\cdot):\mathcal{U}\to\hat{\mathcal{U}}$ represent the mapping from the set of all possible solutions $\mathcal{U}$ to the set of candidate solutions $\hat{\mathcal{U}}$. $U = |\hat{\mathcal{U}}|$ denotes the number of candidate solutions in $\hat{U}$, where each solution corresponds to a set of transmission parameters. In this section, we consider two transmission parameters for each node, namely, the transmission mode $M$ and the transmit power $p$, to optimize the node's transmission rate and signal strength experienced at the receiver. We represent a solution $a \in \hat{U}$ as $\langle M, p \rangle$.

In a multi-user system, adjusting the transmission parameters of individual nodes alone cannot eliminate the strong interference in the overall system \cite{elbatt2004joint}. Therefore, the performance optimization strategy for the UWSNs should incorporate link scheduling mechanisms to coordinate the transmissions of independent users. A reasonable network resource management scheme should first schedule the communication links and then select appropriate transmission parameters for the scheduled nodes, enabling successful transmissions at the receiver. Thus, the action space of nodes consists of link scheduling states and combinations of transmission parameters, denoted as $A=\{wait,\hat{\mathcal{U}}\}$. Here, $|A|=U+1$ represents the number of available actions in the action space. The action \textit{wait} indicates that the node is in a waiting state during the current time slot and does not occupy underwater acoustic network communication resources.

\textcolor{blue}{\textbf{2) Observation $O$.} For any agent $n_i \in \mathcal{N}$, its observation information includes its local information $o_{self}$ and the neighboring node information $o_{neighbor}$. $o_{self}$ comprises: (1) node ID $i$; (2) physical layer status $s_{phy}=\{IDLE,SEND,RECV\}$, indicating whether the node is in idle state, sending state, or receiving state, respectively; (3) node position $\boldsymbol{x}_{i}$; (4) local load information $LI_{i}$. $o_{neighbor}$ includes: (1) neighbor ID $k$, $n_{k} \in \mathcal{N}^{-i}$; (2) neighbor position $\boldsymbol{x}_{k}$; (3) neighbor load information $NL_{k}$.}

\textbf{3) Reward function $R$.} The reward function evaluates the executed joint actions and guide the future behaviors of the nodes. Specifically, TARM evaluates node actions based on the packet reception at the sink node. The reception state of the sink node is denoted as $s_{sink} \in \{recv,idle,conflict\}$, indicating that sink node is receiving a packet, idle, or occurring conflicts among several simultaneous arrivals, respectively. Since the objective of the agent is to maximize the number of successful communications at the sink node, the design of reward can be associated with the reception state of the sink node. Specifically, denoted by $r$, the team reward at each observation duration is described as (\ref{equ:TARM_reward}),
\begin{equation}
r=\left\{
\begin{array}{lcl}
n_{recv},  & & s_{sink}=recv\\
0, & & s_{sink}=idle\\
-n_{conf}, & & s_{sink}=conflict
\end{array} \right.
\label{equ:TARM_reward}
\end{equation}
where $n_{recv}$ represents the number of packets successfully received at the sink, and $n_{conf}$ represents the number of conflicting packets. However, the quantity of successful communications varies when the network traffic differs over a given period of time. To ensure that the designed reward function is applicable to all communication scenarios, we introduce a traffic weight factor $\omega$ to normalize the network traffic of each training episode into a unified range.
\begin{equation}
\omega=\frac{\alpha}{\lambda_{\mathcal{N}}\times \delta_{dur} \times T_{ob}}
\label{equ:TARM_reward}
\end{equation}
Where $\alpha$ is the reward function coefficient and $\lambda_{\mathcal{N}}$ in pkt/s represents the network traffic. In summary, the reward function for TARM is given by equation (\ref{equ:TARM_reward_scale}).
\begin{equation}
r_{TARM}=\omega \times (n_{recv}-n_{conf})\label{equ:TARM_reward_scale}
\end{equation}

\begin{algorithm}
\caption{Training procedure of TARM.}\label{alg:TARM_framework}
Initialize replay buffer $\mathcal{B}$ with size $B$.\\
Initialize mean episode reward list $\bar{\mathbf{R}} \gets \varnothing$.\\
\For{$n_{i} \in \mathcal{N}$}{
Initialize evaluation network $Q_{i}(\boldsymbol{\tau^{t}}, \boldsymbol{a^t}; \theta)$ with random shared parameters $\theta$.\\
Initialize target network $\hat{Q_{i}}(\boldsymbol{\tau^{t}}, \boldsymbol{a^t}; \theta^{-})$ with random shared parameters $\theta^{-}=\theta$.\\
}
\For{$episode=1, \ldots, M$}{
Initialize the environment and receive initial observations $\boldsymbol{o}^{1}$
\For{$t=1, ..., T$}{
\For{$i=1, ..., N$}{
Construct the overhear information matrix according to Algorithm \ref{alg:TARM_confidence}.\\
Concatenate local observation $o_{i}^{t}$.\\
With probability $\epsilon$, randomly select $a_{i}^{t} \in A$; otherwise, select $a_{i}^{t}=\mathop{\arg\max}_{a}{Q_i(\tau_{i}^{t},a_{i})}$.\\
}
Execute joint action $\boldsymbol{a}^{t}=(a_{1}^{t}, ..., a_ {N}^{t})$, receive reward $r$, and obtain new state observations $\boldsymbol{o}^{t+1}=(o_{1}^{t+1}, ..., o_{N}^{t+1})$.\\
Store the transition $(\boldsymbol{o}^{t}, \boldsymbol{a}^{t}, r, \boldsymbol{o}^{t+1})$ in $\mathcal{B}$
}
Randomly sample $b$ transitions $\mathcal{B}$, and train the evaluation network $\theta$ with the mini-batch data.\\
%Train the evaluation network $\theta$ with the mini-batch data.\\
Update $\theta^{-}=\theta$ every $\mathcal{C}$ episode, and calculate the mean episode reward $\bar{R}(\theta_{episode})$.\\
$\bar{\mathbf{R}}[episode-1]=(\bar{R}(\theta_{episode}),\theta_{episode})$.\\
}
\Return $\theta=\mathop{\arg\max}_{\bar{\mathbf{R}}}\bar{R}(\theta_{episode})$
\end{algorithm}

TARM is implemented with the deep recurrent Q-network (DRQN) \cite{hausknecht2015deep} and the traffic load-aware mechanism proposed in this paper. In distributed MAS such as the UWSNs considered in this paper, the agent does not have full observability of the state of all other agents in the system, and can not access to the global reward signals considering the computation and transmission overhead \cite{zhang2011scalingPartial}, which is necessary for model training. To this end, the centralized training with decentralized execution (CTDE) paradigm \cite{oliehoek2008optimal} is adopted to train TARM. Specifically, during the training phase, a virtual central controller is instantiated on the sink node to collect the state information of the underwater nodes and the communications, which then evaluates the value of joint actions performed by the agents and distributes the reward signal to each agent. In the underwater system considered in this paper, all agents work cooperatively to optimize the network performance. 

One of the promising ways to exploit the CTDE paradigm is value decomposition network (VDN) \cite{sunehag2018value}, which learns a decentralized utility function for each agent and uses a mixing network to combine these local utilities into a team reward \cite{wang2020roma}. Note that TARM is not limited to any team reward mixing network, only if it can reflect the influence of the joint actions on the environment. The training process of TARM is illustrated in Algorithm \ref{alg:TARM_framework}, which facilitates the decision-making ability of the TARM agents in choosing optimal actions with its partial observations.

\begin{figure*}[htbp]
\centerline{\includegraphics[width=5.5in]{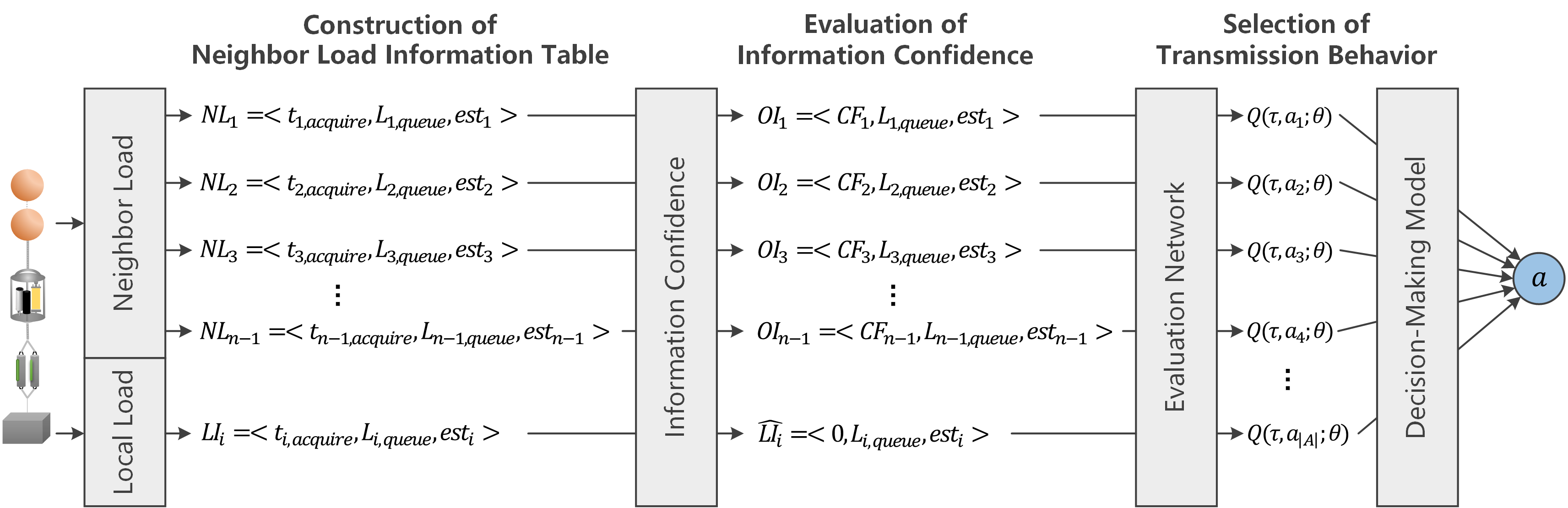}}
\caption{\textcolor{blue}{Decision process of TARM node.}}
\label{fig:TARM_decision}
\end{figure*}

During the decentralized execution phase, the virtual central controller on the sink node is removed, and the transmitters only retain their trained TARM models to select actions based on the local observations, without receiving feedback signals from the sink node or negotiating with neighboring nodes. \textcolor{blue}{The decentralized decision-making process, including the construction of the neighbor load information table, evaluation of information confidence, and selection of transmission behaviors via a behavior policy, is illustrated in Fig. \ref{fig:TARM_decision}. Specifically, the behavior policy during the execution phase utilizes the trained RL-based decision-making model.}

During the construction phase of the neighbor load information table $T_{NL}$, each node maintains the latest neighbor node load information. When a neighbor node information $NL_{j}=\langle t_{j,acquire}, L_{j,queue}, est_{j}\rangle$ is detected, the table is checked for existing records from that node. If there is a record, $NL_{j}$ is updated; otherwise, the current information is added to the table. The construction phase of $T_{NL}$ spans the entire lifetime of the network. 

In each decision time slot, the node enters the information confidence evaluation stage, in which the node assigns confidence to each record in $T_{NL}$ based on the difference between the information acquisition time and the current decision time. This measures the reliability of the information for subsequent decision-making stages. The acquisition time $t_{j,acquire}$ in $NL_{j}$ is replaced with the information confidence $CF_{j}$. Hence, the information used as input to the decision-making network is $OI_{j}=\langle CF_{j}, L_{j,queue}, est_{j}\rangle$, and $\mathbf{OI}=\{OI_{j}\}_{\mathcal{N}^{-i}}$ represents the overhear information matrix. It should be noted that for each node, the acquisition time of its local information is always synchronized with the decision time, so the confidence of local information used for decision-making is set to 0.

The decision network of the TARM agent takes each overhear information $OL_{j}$ and local confident load information $\hat{LI}_{i}=\langle 0, L_{i,queue}, est_{i}\rangle$ as inputs. Through the nonlinear transformation of the deep decision network, it outputs the Q-value corresponding to each optional action in the action set $A$, i.e., $Q(\tau,a_{i};\theta)$. Here, $\theta$ represents the parameters of the deep decision network that have been trained during the centralized training phase. Finally, based on the Q-values of each action, the node selects the action with the maximum Q-value for transmission. It can be seen that during the execution phase, TARM makes real-time decisions by selecting transmission parameters based on the overhear information and local information in an end-to-end manner. \textcolor{blue}{Specifically, due to the unreliable nature of the underwater acoustic channel, the observation information, including overhearing information from neighboring nodes, may only be available periodically. In light of this practical issue, TARM schedules the communication links as in the previous slot. Once the observation information is acquired, TARM adjusts the transmission schedule to accommodate the current communication and traffic loads.}

\section{Performance Evaluation}\label{sec:TARM_eva}

In this section, we evaluate the performance of TARM in single-hop UWSNs for various network settings such as different traffic load and packet size of underwater nodes. Our simulations address the following questions:
\begin{enumerate}
\item How does TARM adapt as the communication demand changes ($\S$\ref{sec:TARM_result_load})?
\item How does TARM perform as the packet collision probability gradually increases ($\S$\ref{sec:TARM_result_size})?
\item How does each of the key components of TARM contribute to its performance ($\S$\ref{sec:TARM_result_comp})?
\end{enumerate}

\subsection{Evaluation Setup}\label{sec:TARM_para}

Considering a single-hop UWSN with three transmitters send and a sink node. where the transmitters are randomly deployed at a distance of 3,000-5,000 m from the sink node. \textcolor{blue}{The underwater nodes operate at 24 kHz carrier frequency with 6 kHz bandwidth.} The observation duration $T_{ob}$ is set to 200 time slots, and duration $\delta_{dur}$ of each slot is 1 second. Other simulation parameters are summarized in Table \ref{tab:TARM_paraset}.
\begin{table}[htp]
\caption{Parameters.}
\begin{center}
\begin{tabular}{ll}
\hline
Parameter&Value\\
\hline
Range of transmit power $p_{min}$, $p_{max}$&0 W, 40 W\\
Receiving power $p_{recv}$&0.395 W\\
Idle power $p_{idle}$&0.001 W\\
Transmission mode $\mathcal{M}$&$\{1,2,3,4,5\}$\\
Transducer efficiency $\eta_{0}$&0.6\\
Population size $N_{pop}$&500\\
Number of generation $T_{gen}$&500\\
\hline
\end{tabular}
\end{center}
\label{tab:TARM_paraset}
\end{table}

In this section, we adopt the five transmission modes available in the OFDM modem from AquaSeNT \cite{wan2013field}, as shown in Table \ref{tab:TARM_OFDM}. The bandwidth $B$ of the modem is 6000 Hz, and the number of subcarriers $K$ is 1024, resulting in a symbol duration $\delta_{block}$ of 170.7 milliseconds (ms). Among the 1024 subcarriers, 256 pilot subcarriers are used for channel estimation, and there are also 96 null subcarriers for Doppler compensation at the receiver. Therefore, the number of subcarriers used for data transmission is 672, so that each codeword can be accommodated in one OFDM symbol regardless of the transmission mode adopted.
\begin{table}[htp]
\caption{Transmission mode of the adopted acoustic modem.}
\begin{center}
\begin{tabular}{p{0.8cm}<{\centering}p{1.2cm}<{\centering}p{1.6cm}<{\centering}p{1.1cm}<{\centering}p{1.2cm}<{\centering}}
\hline
Mode&Code rate&Modulation scheme&Payload (Byte)&Rate (kb/s)\\
\hline
1&1/2&BPSK&38&1.38\\
2&1/2&QPSK&80&2.90\\
3&3/4&QPSK&122&4.42\\
4&1/2&16QAM&164&5.94\\
5&3/4&16QAM&248&8.99\\
\hline
\end{tabular}
\end{center}
\label{tab:TARM_OFDM}
\end{table}

Depending on the adopted code rates and modulation schemes, different transmission modes exhibit distinct transmission rates and statistical thresholds necessary for effective communication. The statistical thresholds for the five transmission mode are set to 3.8, 5.0, 7.4, 9.2, and 12.2 dB, which are determined according to sea trial results \cite{wan2014adaptive}. 
\begin{figure}[htbp]
\centerline{\includegraphics[width=3.5in]{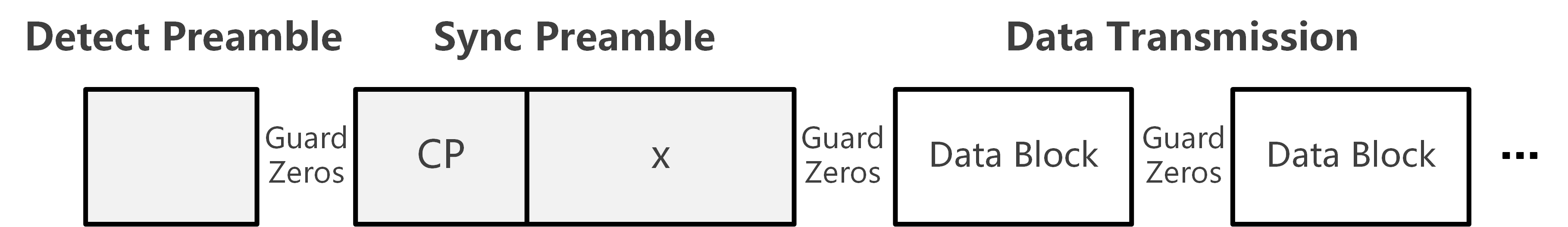}}
\caption{Structure of the OFDM data packet.}
\label{fig:TARM_dataPacket}
\end{figure}

The acoustic modem employed in this paper incorporates a data packet structure as depicted in Fig. \ref{fig:TARM_dataPacket}. This structure comprises three key components: a detection preamble for triggering the receiver to process the data packet, a synchronization preamble for synchronizing and conveying control information (such as transmission mode and length of guard zeros for the following data blocks), and several data blocks. The number of data blocks $n_{block}$ is determined by data length $L_{d}$ and the payload per block $L_{b}$. It is important to note that the payload per block varies depending on the transmission mode adopted. During the evaluation phase, specific durations are established for the transmission and reception processes. The duration for transmitting the two preambles $\delta_{pre}$ is set to 500 ms. Additionally, a guard interval of 50 ms, denoted as $\delta_{guard}$, is included between consecutive data blocks. The exact value of the propagation delay, denoted as $\delta^{prop}$, is obtained through the BELLHOP ray tracing algorithm. \textcolor{blue}{The sea bottom parameters, including density, attenuation, and sound speed are set to 1.90 g/cm$^{3}$, 0.46 dB/$\lambda$, respectively \cite{huang2020adaptive}.}

The construction of the evaluation and target networks involves the utilization of two fully connected (FC) layers and a GRU within the framework. Specifically, the first FC layer consists of 64 hidden units, followed by a GRU layer with 64 hidden units, and a Rectified Linear Unit (ReLU) to abstract the observations. Subsequently, the second FC layer comprises seven hidden units that generate Q-values for each possible action. In terms of the training process, an experience buffer with a capacity of 10,000 is employed, along with a mini-batch size of 32. The RMSprop optimizer with a learning rate of 0.0005 is utilized for all networks \cite{zou2019sufficient}. The model was trained on 200,000 episodes. Every 200 episodes, an evaluation is conducted by executing the model 20 times and calculating the mean episode rewards. The optimal model is determined based on the highest average reward achieved. To strike a balance between exploration and exploitation during training, an $\varepsilon$-greedy behavior policy is adopted. Initially, $\varepsilon$ decreases linearly from 1 to 0.05 over the first 100,000 episodes, and then stay constant at $\varepsilon$=$0.05$ for the remaining 100,000 episodes. These settings enable the agents to thoroughly explore the environment in the early stages while gradually converging in the later stages, thus facilitating the collaborative maximization of long-term returns \cite{mnih2015human}. The discount factor $\gamma$ is set to 0.99 to appropriately weigh immediate and future rewards in the decision-making process.

All training and simulations in this research are conducted in Python. The implementation of DRL is facilitated by the Pytorch library \cite{paszke2019pytorch}. The simulations are executed on a computer system equipped with a GeForce RTX 3070Ti GPU, an Intel i9-12900H CPU@ 2.5GHz, and 32GB of RAM.

\subsection{Baseline Methods and Evaluation Metrics}\label{sec:TARM_baseline}

To validate the effectiveness of the proposed TARM strategy, we compare TARM's performance to that of the following baseline methods. Table \ref{tab:TARM_baseline} presents a comparison between baseline methods and TARM in relation to their optimization objectives and adopted mechanisms. The optimization objectives, denoted as $f_1$, $f_2$, and $f_3$, represent the optimization of network throughput, transmission time, and energy consumption, respectively.
\begin{enumerate}[1)]
\item \textbf{Slotted-Aloha \cite{chirdchoo2007aloha}.} The node schedules the communication links based on the slotted-Aloha mechanism to transmit data packets with maximum transmit power and minimum transmission rate, without considering network performance or other communication constraints.
\item \textbf{Slotted-Aloha (Min $f_3$).} The node employs the slotted-Aloha mechanism to schedule communication links and selects transmission parameters with the optimization objective of minimizing energy consumption, as outlined in the third optimization objective of TARM.
\item \textbf{Slotted-Aloha (Min $f_2$).} The node utilizes the slotted-Aloha mechanism to schedule communication links and selects transmission parameters with the optimization objective of minimizing packet transmission time, as stated in the second optimization objective of TARM.
\item \textbf{NF-TDMA \cite{diamant2016leveraging}.} The underwater nodes are organized under the TDMA scheme, with time slots allocated using the near-far effect. NF-TDMA is implemented to enhance network throughput and reduce transmission delay.
\item \textbf{DR-DLMA \cite{ye2020deep}.} The nodes judiciously utilize the available time slots resulted from propagation delays or not used by other nodes to maximize the network throughput based on a deep reinforcement learning algorithm.
\end{enumerate}

\begin{table}[htp]
\caption{Comparisons of baseline methods and TARM.}
\begin{center}
\begin{tabular}{lp{0.3cm}<{\centering}p{0.3cm}<{\centering}p{0.3cm}<{\centering}p{1.3cm}<{\centering}p{1cm}<{\centering}}
\hline
Method&$f_{1}$&$f_{2}$&$f_{3}$&Propagation delay&Traffic aware\\
\hline
Slotted-Aloha&/&/&/&/&/\\
Slotted-Aloha (Min $f_3$)&/&/&$\surd$&/&/\\
Slotted-Aloha (Min $f_2$)&/&$\surd$&/&/&/\\
NF-TDMA&$\surd$&$\surd$&/&$\surd$&/\\
DR-DLMA&$\surd$&$\surd$&/&$\surd$&/\\
TARM \textbf{[ours]}&$\surd$&$\surd$&$\surd$&$\surd$&$\surd$\\
\hline
\end{tabular}
\end{center}
\label{tab:TARM_baseline}
\end{table}

\begin{figure*}[htbp]
\centering
\subfigure[Throughput]{
\begin{minipage}[t]{0.3\linewidth}
\centering
\includegraphics[width=2.2in]{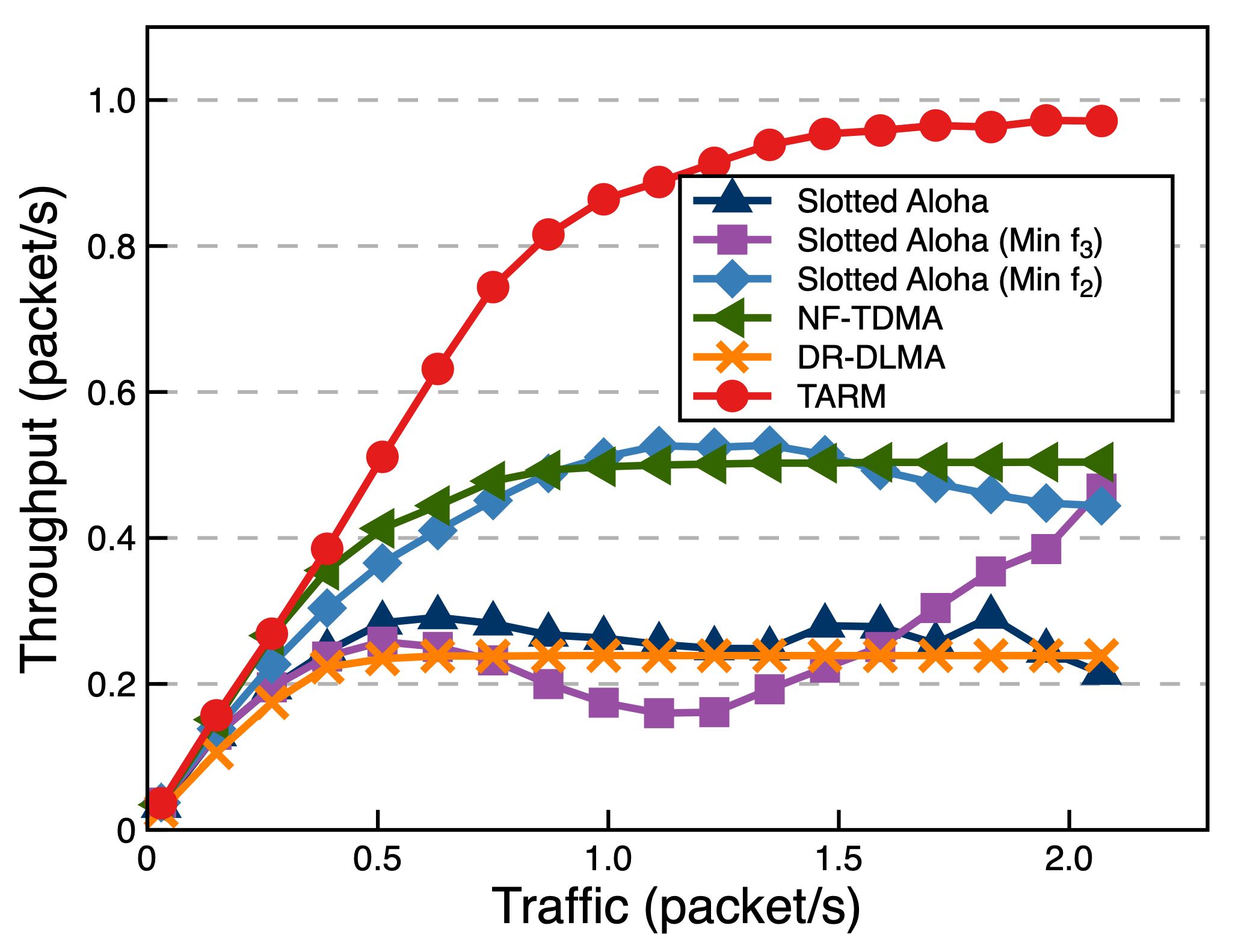}
\label{fig:TARM_traffic_throughput}
\end{minipage}
}%
\subfigure[End-to-end delay]{
\begin{minipage}[t]{0.3\linewidth}
\centering
\includegraphics[width=2.2in]{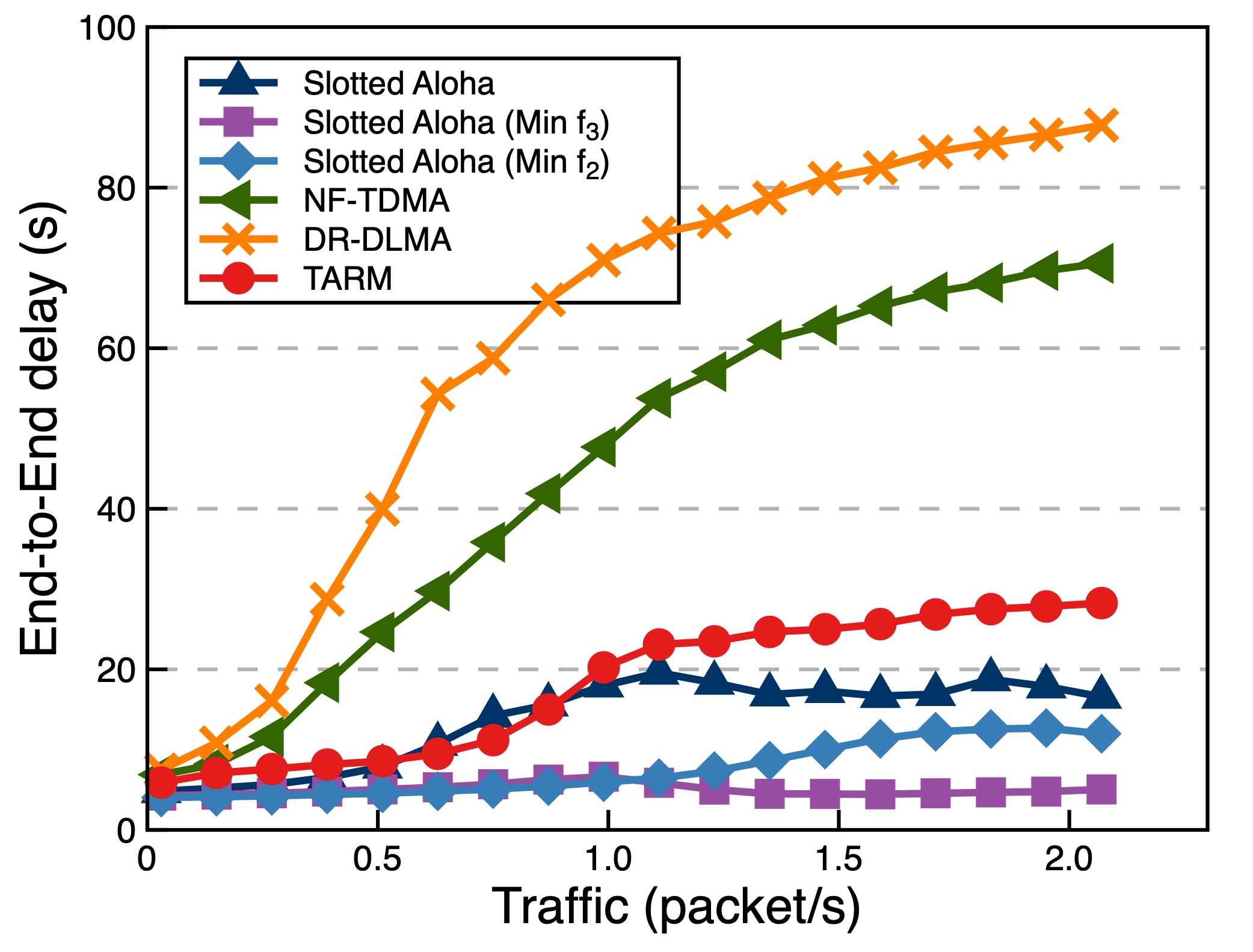}
\label{fig:TARM_traffic_delay}
\end{minipage}
}%
\subfigure[Average energy consumption]{
\begin{minipage}[t]{0.3\linewidth}
\centering
\includegraphics[width=2.2in]{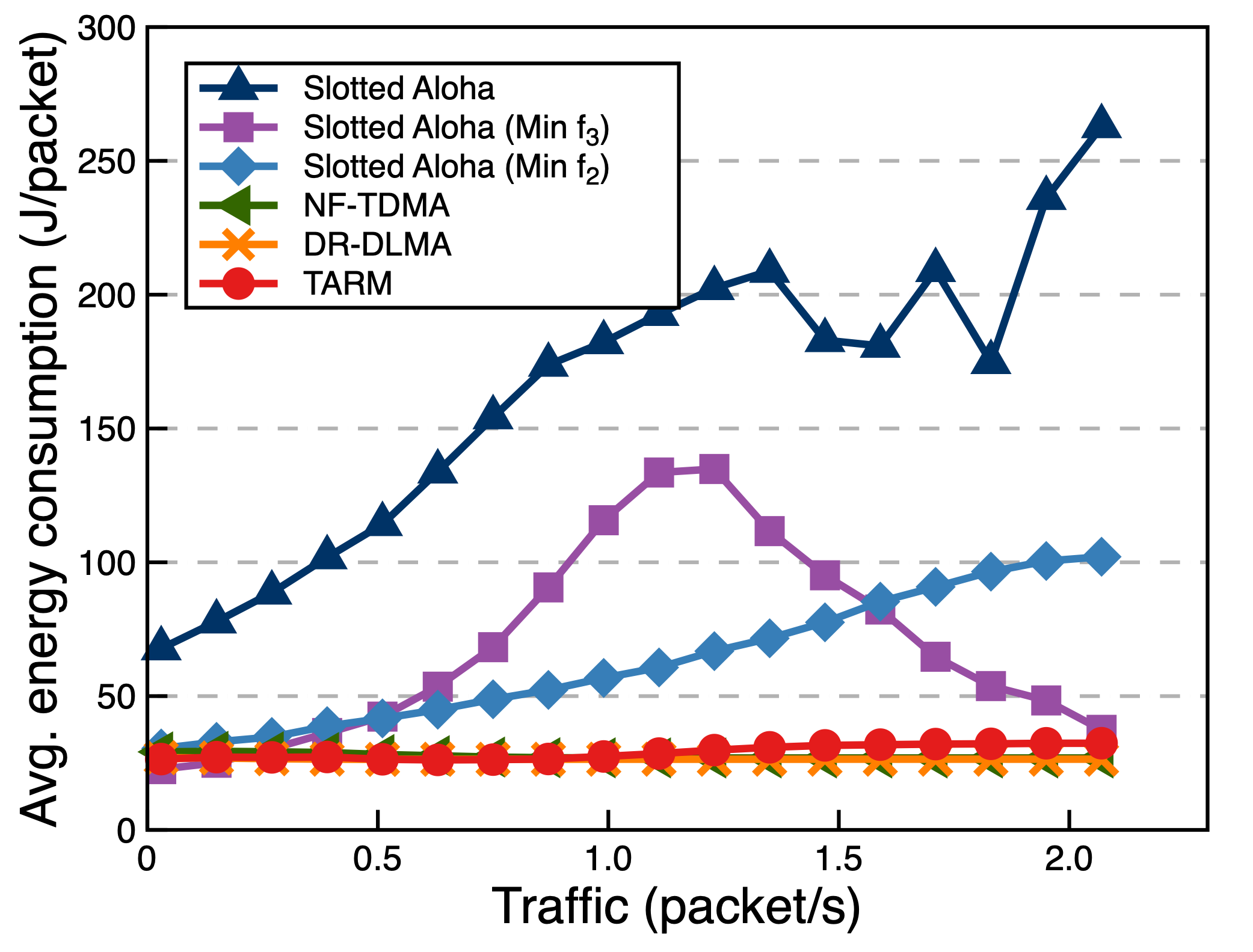}
\label{fig:TARM_traffic_energy}
\end{minipage}
}%
\\
\subfigure[Delivery ratio]{
\begin{minipage}[t]{0.3\linewidth}
\centering
\includegraphics[width=2.2in]{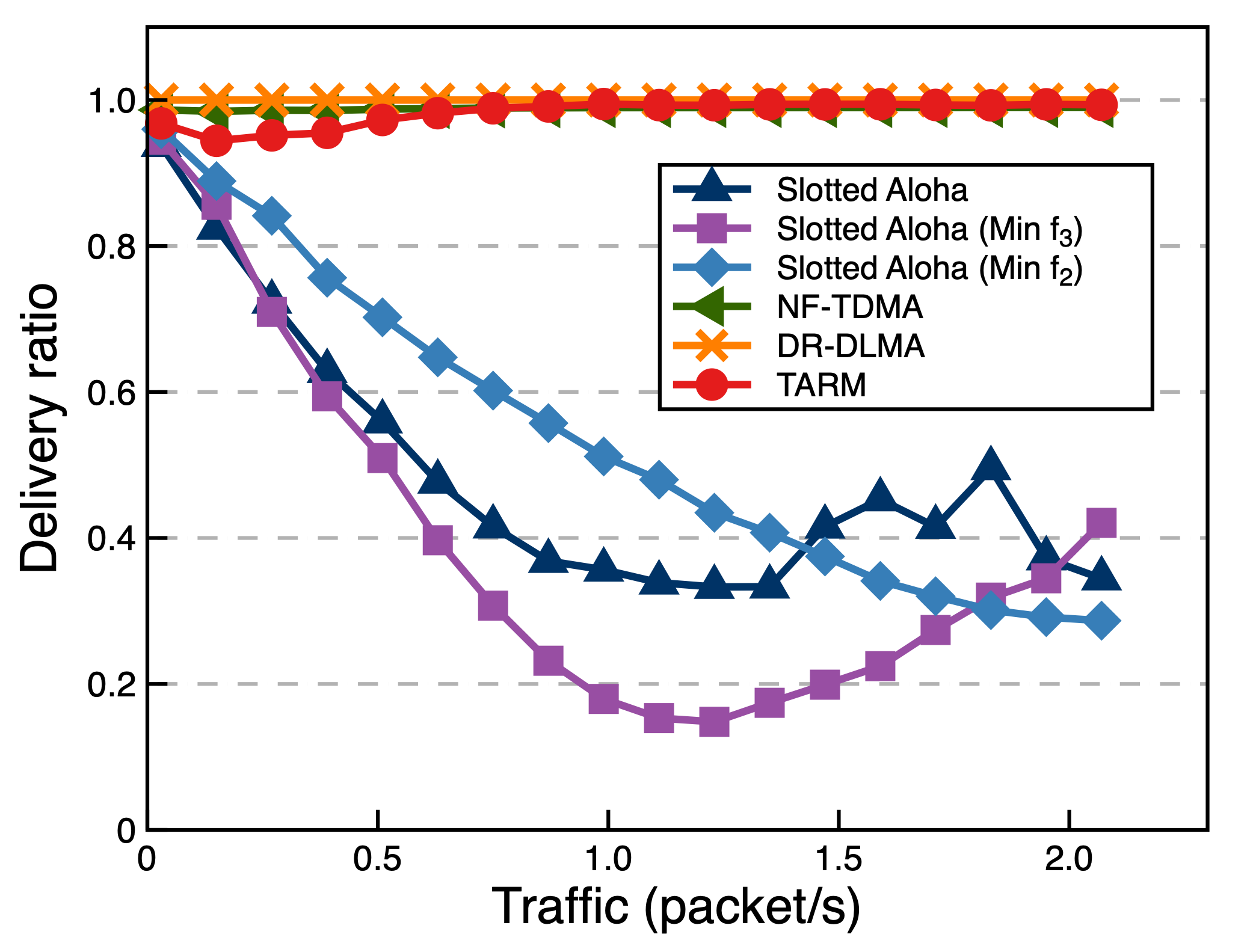}
\label{fig:TARM_traffic_delivery}
\end{minipage}
}%
\subfigure[Channel utilization]{
\begin{minipage}[t]{0.3\linewidth}
\centering
\includegraphics[width=2.2in]{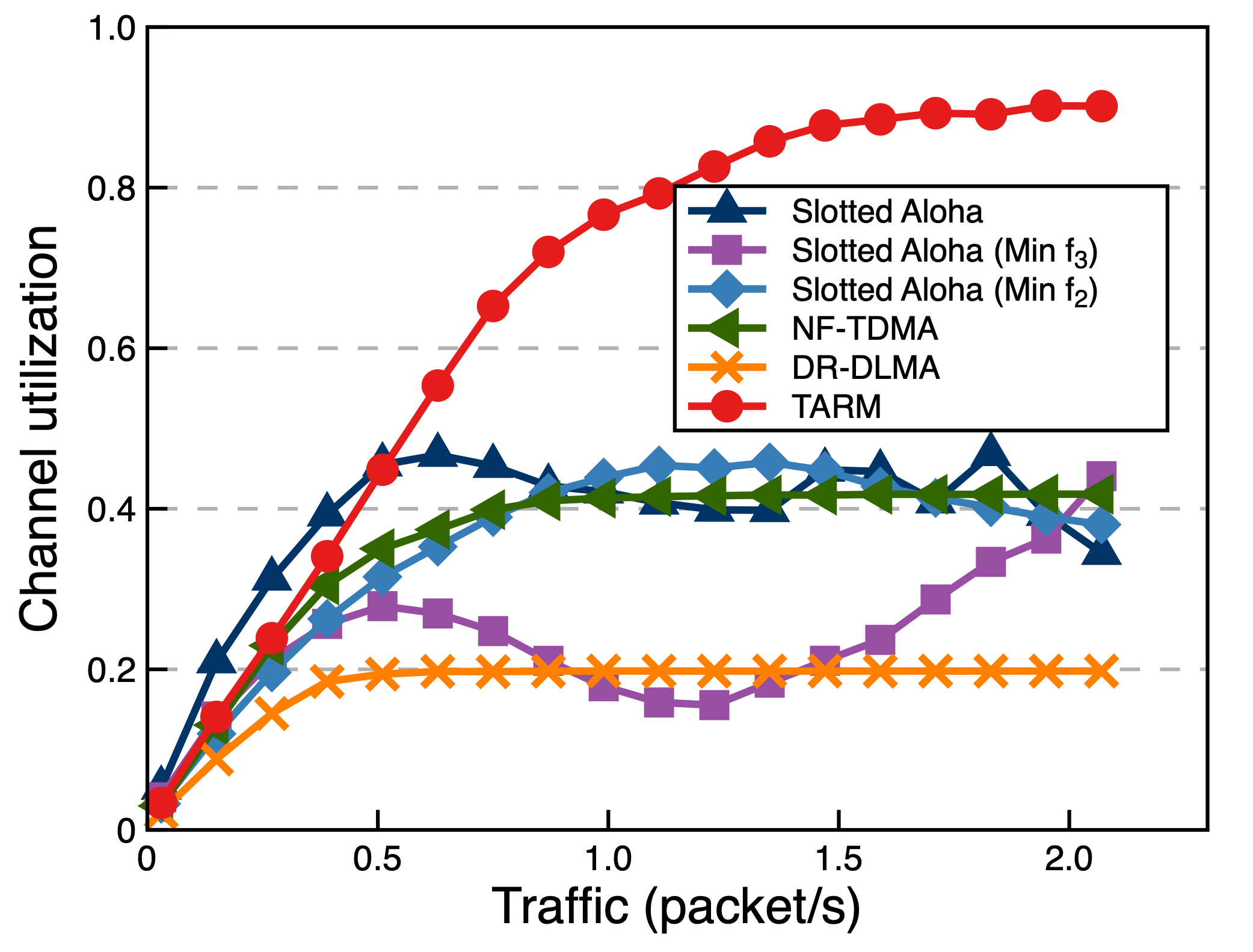}
\label{fig:TARM_traffic_utilization}
\end{minipage}
}%
\centering
\caption{Impact of traffic load on network performance.}
\label{fig:TARM_traffic}
\end{figure*}

The evaluation metrics include: (a) network throughput $C_{pkt}=re_{\mathcal{N}}(T_{ob})/T_{ob}$, where $re_{\mathcal{N}}(T_{ob})$ is defined as (\ref{equ:TARM_N_success}); (b) average end-to-end delay $\bar{\delta}_{end}=(\sum_{j=1}^{re_{\mathcal{N}}(T_{ob})}\delta_{end}^{j})/re_{\mathcal{N}}(T_{ob})$, where $\delta_{end}^{j}$ is defined as the duration from the packet is generated to it is successfully received at the receiver; (c) average energy consumption $\bar{E}_{pkt}=(\sum_{i=1}^{N}\sum_{j=1}^{s_{i,m}(T_{ob})}e_{i,m}^{j})/re_{\mathcal{N}}(T_{ob})$, (d) delivery ratio $D_{\mathcal{N}}=re_{\mathcal{N}}(T_{ob})/\sum_{i=1}^{N}s_{i,m}(T_{ob})$, and (e) channel utilization $U=(\sum_{j=1}^{re_{\mathcal{N}}(T_{ob})}\delta_{j}^{tx})/T_{ob}$.

\subsection{Impact of Traffic Load on Network Performance}\label{sec:TARM_result_load}

In this section, we investigate the performance of the Traffic Load-Aware Resource Management (TARM) strategy and baseline methods under increasing network traffic load. The network traffic, denoted as $\lambda_{\mathcal{N}}$, serves as an indicator of the communication demand. Each node generate its own traffic based on a Poisson process with a density parameter of $\lambda_{\mathcal{N}}$. The value of $\lambda_{\mathcal{N}}$ progressively increases from 0.03 to 2.07 packets per second with a step size of 0.12, capturing diverse communication requirements as depicted in Figs. \ref{fig:TARM_traffic_throughput}-\ref{fig:TARM_traffic_utilization}. During the evaluations, all baseline methods employ a fixed packet length of 190 bytes. However, for the implementation of the traffic load-aware mechanism in the TARM approach, an additional 10 bytes are required to accommodate the load information (i.e., information acquisition time, node queue length, and local traffic estimation), resulting in a total packet length of 200 bytes. Each simulation experiment consists of 200 time slots, and the evaluation results presented are an average of 100 independent executions.
\begin{figure*}[htbp]
\centering
\subfigure[Throughput]{
\begin{minipage}[t]{0.3\linewidth}
\centering
\includegraphics[width=2.2in]{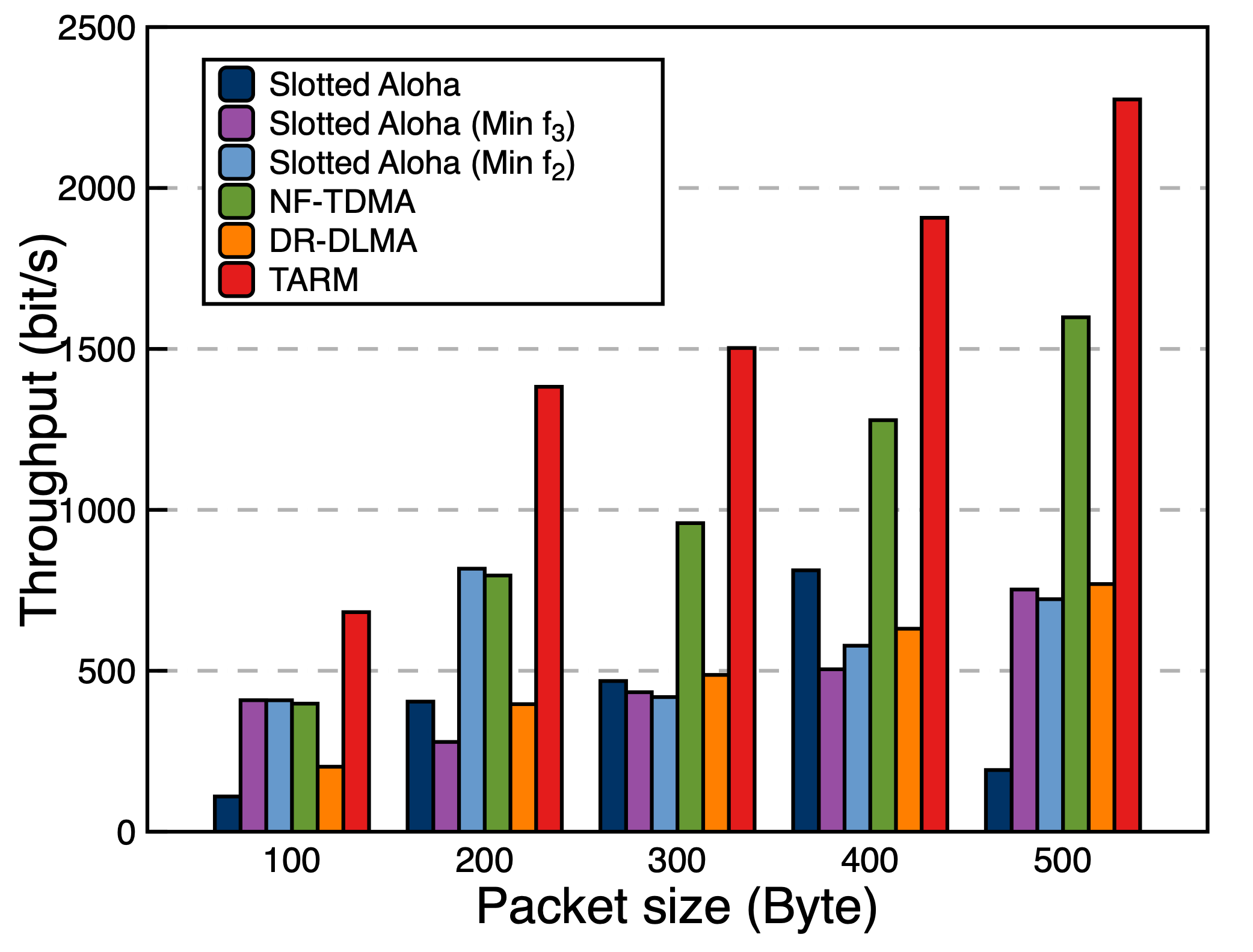}
\label{fig:TARM_packet_throughput}
\end{minipage}
}%
\subfigure[End-to-end delay]{
\begin{minipage}[t]{0.3\linewidth}
\centering
\includegraphics[width=2.2in]{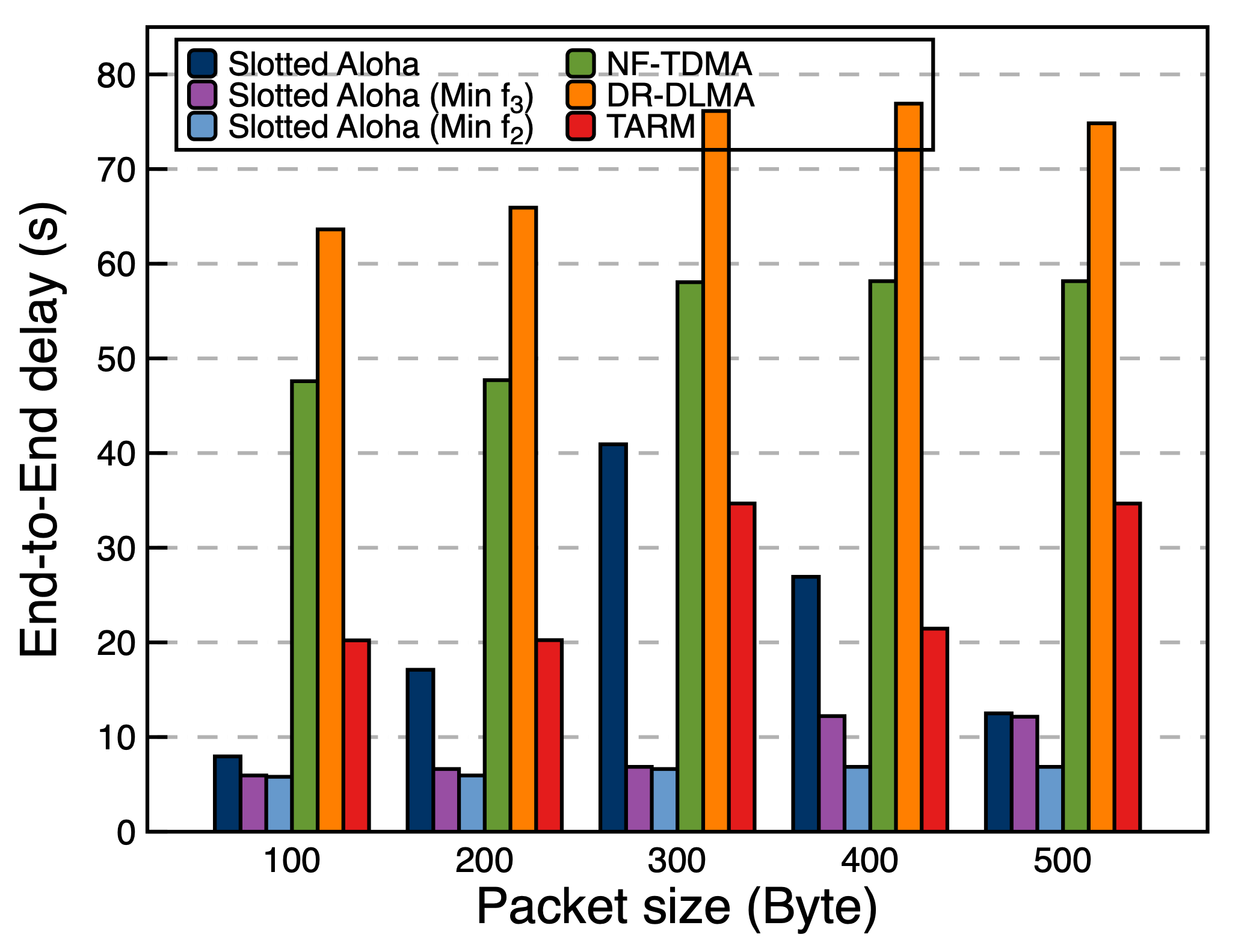}
\label{fig:TARM_packet_delay}
\end{minipage}
}%
\subfigure[Average energy consumption]{
\begin{minipage}[t]{0.3\linewidth}
\centering
\includegraphics[width=2.2in]{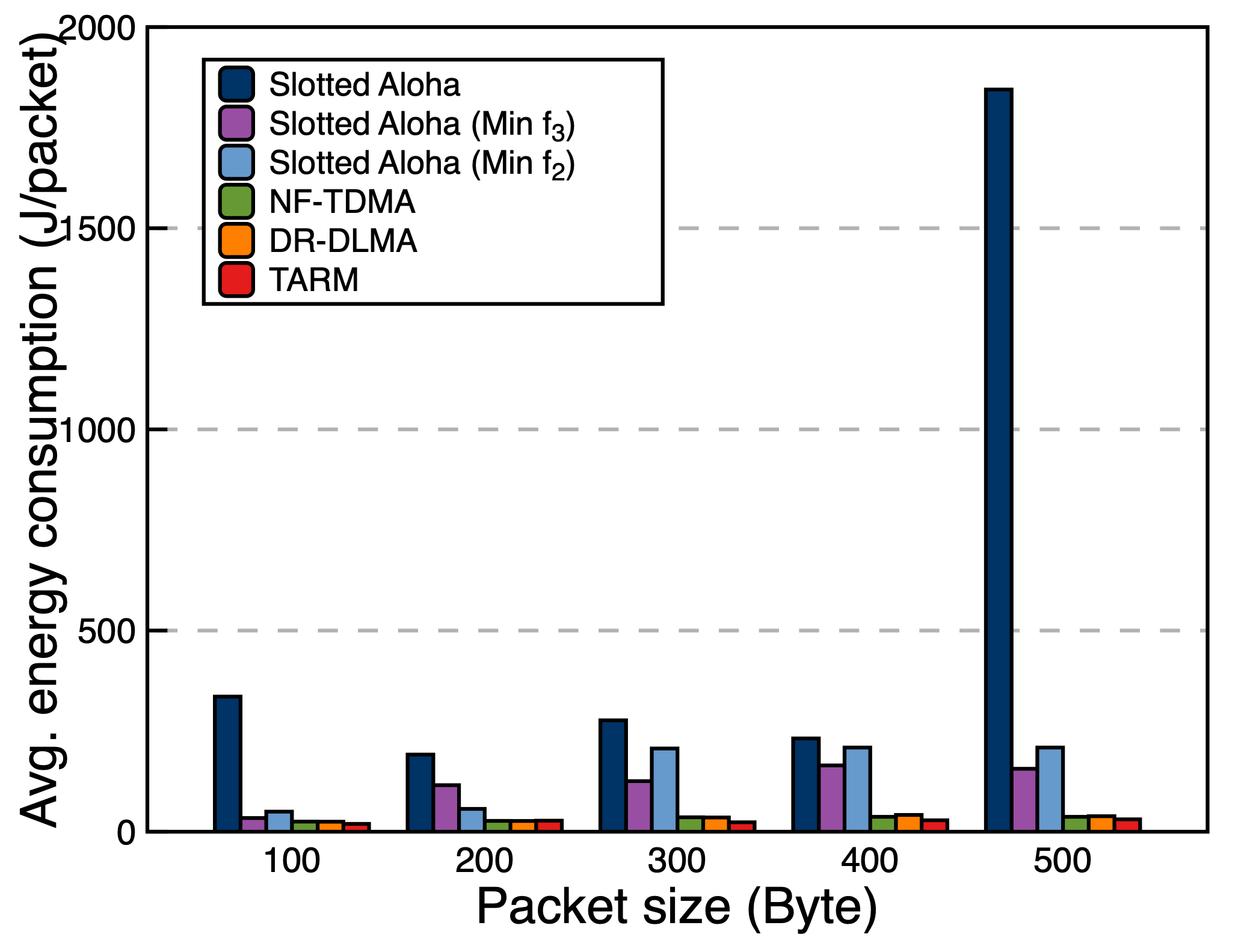}
\label{fig:TARM_packet_energy}
\end{minipage}
}%
\\
\subfigure[Delivery ratio]{
\begin{minipage}[t]{0.3\linewidth}
\centering
\includegraphics[width=2.2in]{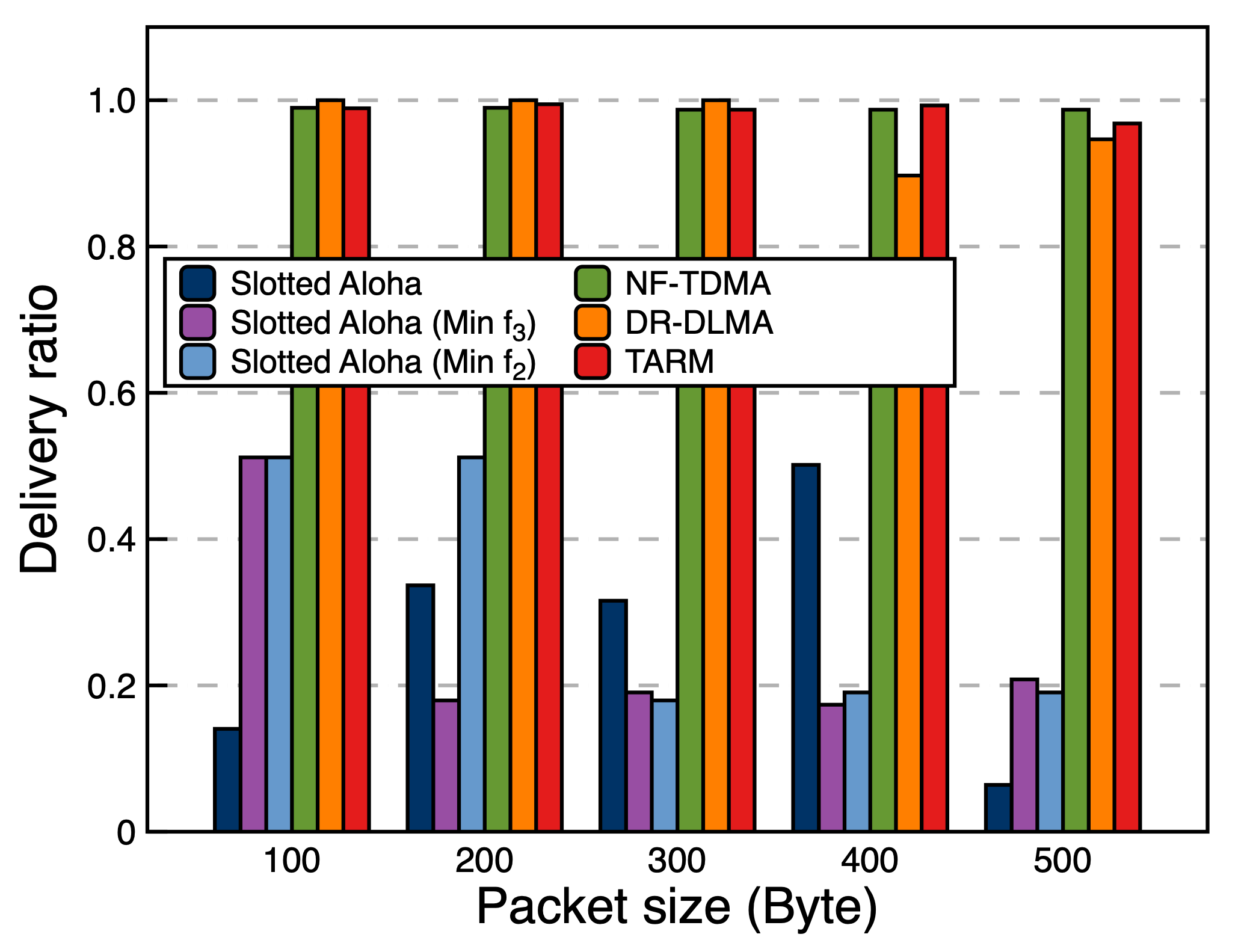}
\label{fig:TARM_packet_delivery}
\end{minipage}
}%
\subfigure[Channel utilization]{
\begin{minipage}[t]{0.3\linewidth}
\centering
\includegraphics[width=2.2in]{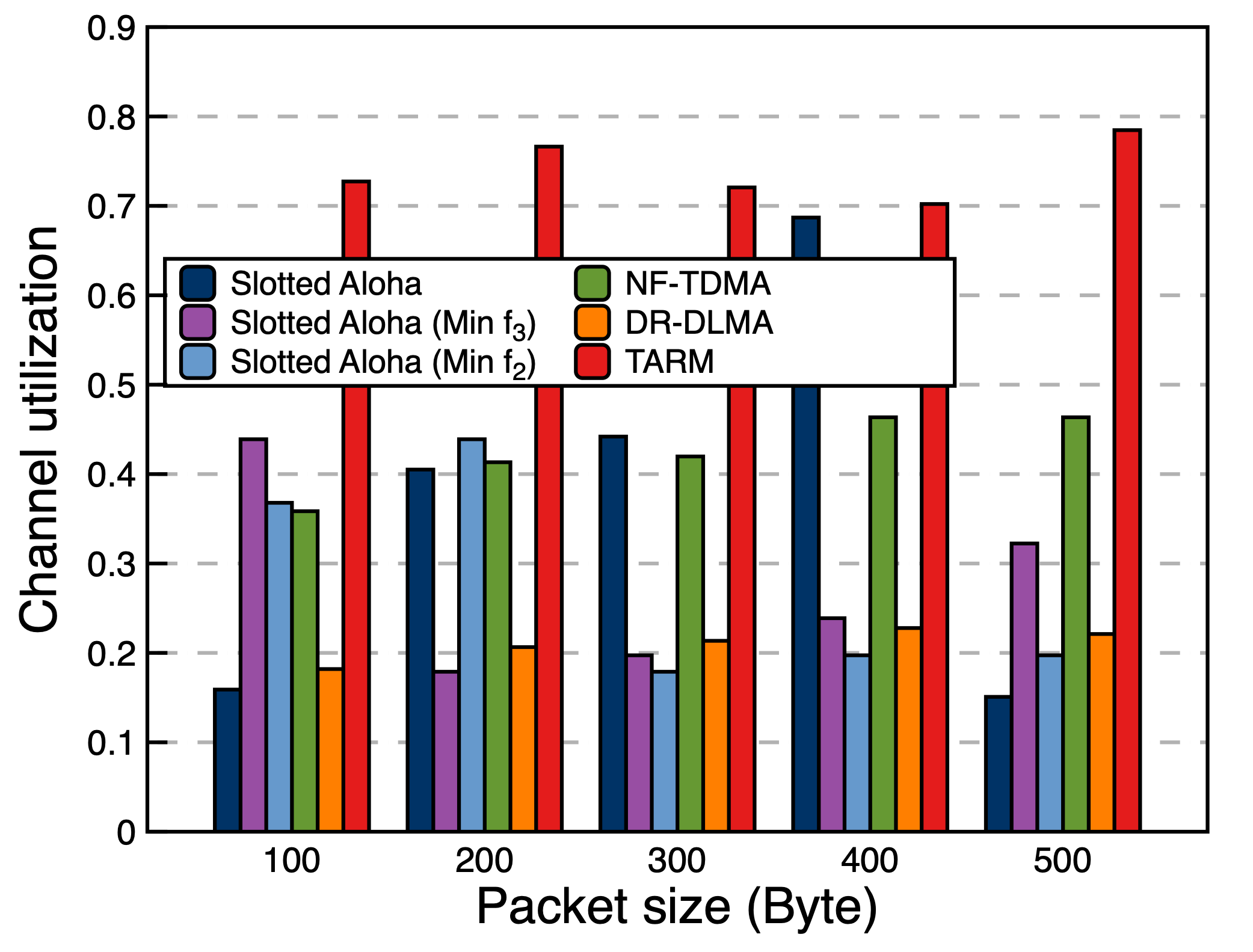}
\label{fig:TARM_packet_utilization}
\end{minipage}
}%
\centering
\caption{Impact of data packet size on network performance.}
\label{fig:TARM_packet}
\end{figure*}

As illustrated in Fig. \ref{fig:TARM_traffic}, the network traffic directly impacts the performance of TARM, particularly in terms of throughput, channel utilization, delivery ratio, and average packet energy consumption. Under increasing network traffic conditions, TARM exhibits notable improvements. When the network traffic load is low ($\lambda_{\mathcal{N}} < 0.27$ pkt/s), all methods exhibit similar throughputs. This phenomenon can be attributed to the presence of abundant communication resources during periods of low network loads, thereby facilitating the optimal utilization of these resources by all methods. As a result, the disparities in throughput observed among these methods are negligible in this context. However, as $\lambda_{\mathcal{N}}$ increases, the throughput differences among the methods gradually increase. Notably, TARM demonstrates superior performance under high load scenarios where link scheduling and resource management strategies play a crucial role in network performance, especially when communication resources become constrained. The throughput of TARM experiences rapid growth when $\lambda_{\mathcal{N}}<0.75$ pkt/s, after which it gradually slows down and stabilizes as $\lambda_{\mathcal{N}}$ approaches 1.71 pkt/s. Specifically, at $\lambda_{\mathcal{N}}=0.75$ pkt/s, TARM achieves a throughput increase ranging from 55.6\% to 220.2\% compared to the baseline methods, at $\lambda_{\mathcal{N}}=2.07$ pkt/s, TARM achieves a remarkable throughput increase ranging from 92.8\% to 351.6\% compared to the baseline methods. These results highlight the effectiveness of the TARM approach in enhancing network performance under high traffic load conditions.

The performance evaluation of Slotted Aloha provides insights into network behavior when lacking any link scheduling or resource management strategy. To ensure reliable communication, nodes utilizing Slotted Aloha adopt the minimum transmission mode ($M_{i \in \mathcal{N}}=1$) and maximum transmission power ($p_{i \in \mathcal{N}}=p_{max}$). Within the set of candidate solutions obtained through the SSO algorithm, Slotted Aloha (Min $f_3$) selects the solution that minimizes energy consumption, while Slotted Aloha (Min $f_2$) opts for the solution with the lowest transmission delay. Ideally, Slotted Aloha (Min $f_3$) would exhibit the minimum energy consumption, and Slotted Aloha (Min $f_2$) would demonstrate the shortest end-to-end delay. However, these expectations do not align with the simulation results. Regarding network throughput, Slotted Aloha outperforms the other two resource management policies in communication scenarios with $\lambda_{\mathcal{N}}<1.71$ pkt/s. This suggests that suboptimal resource management solutions struggle to effectively optimize network performance in the majority of cases.

When $\lambda_{\mathcal{N}}$ exceeds 0.39, the network guided by DR-DLMA gradually approaches its maximum achievable network throughput. This is because DR-DLMA does not exploit the long propagation delay characteristic of acoustic channels and strictly adheres to the link scheduling principles of Slotted TDMA, where at most one node can access to the acoustic channel during a given transmission slot. The primary advantage of DR-DLMA lies in its ability to completely avoid communication collisions among nodes, leading to satisfied delivery ratio and average energy consumption. However, the drawback of DR-DLMA becomes apparent when examining the end-to-end delay metric. As DR-DLMA nodes have fewer opportunities for transmission compared to other methods, their delivery delay increases abruptly as the network traffic grows. Furthermore, the channel utilization achieved by DR-DLMA is inferior to that of its counterparts.

The aforementioned observations validate the effectiveness of reasonable link scheduling and resource management strategies in optimizing network performance, particularly under high network traffic conditions. Subsequently, we evaluate the performance of TARM against the baseline methods in networks characterized by varying levels of conflict probabilities, i.e., different packet lengths.

\subsection{Impact of Data Packet Size on Network Performance}\label{sec:TARM_result_size}

The length of data packets in UWSNs has great influence on packet collision probability and channel utilization. Longer packets allow for more efficient transmission of larger amounts of data but also increase the risks of packet collisions. On the other hand, shorter packets reduce the probability of collisions but result in inefficient communication due to higher overhead caused by frequent packet headers. This section analyzes the performance of different methods under varying packet lengths, as depicted in Figs. \ref{fig:TARM_packet_throughput}-\ref{fig:TARM_packet_utilization}. TARM exhibits notable advantages concerning network throughput and energy consumption. It can provide reliable communications for UWSNs with a delivery ratio close to 1. Given that packet length directly influences the number of transmitted bits, this section evaluates the throughput of different techniques in terms of bit rate (bit/s). The network traffic remains constant at 0.99 pkt/s across all scenarios. Each simulation experiment consists of 200 time slots, and the evaluation results are the average values from 100 simulation experiments.
\begin{figure}[htbp]
\centerline{\includegraphics[width=3in]{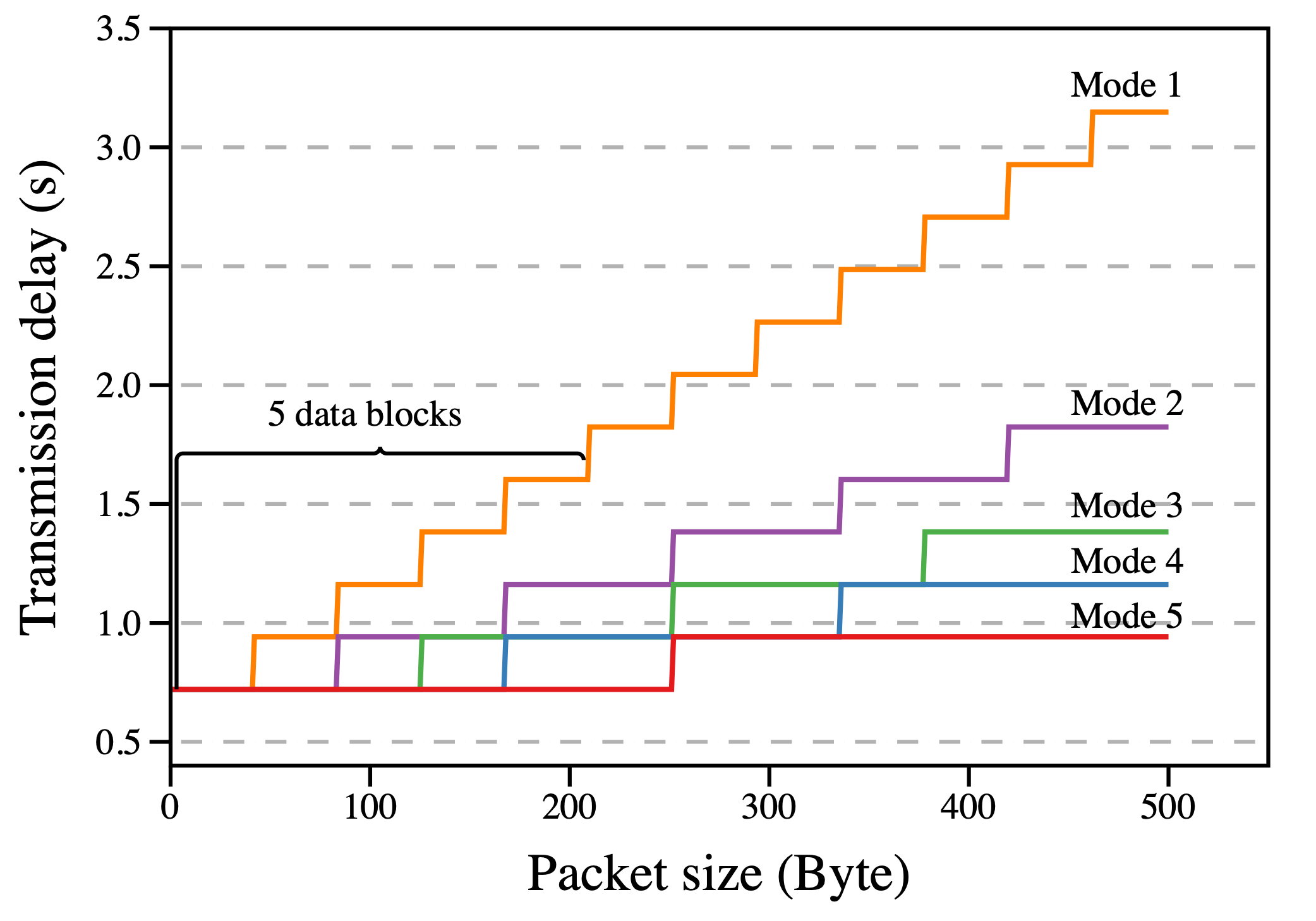}}
\caption{Transmission delay of the OFDM packet.}
\label{fig:delay_size}
\end{figure}
\begin{figure*}[htbp]
\centering
\subfigure[Throughput]{
\begin{minipage}[t]{0.3\linewidth}
\centering
\includegraphics[width=2.2in]{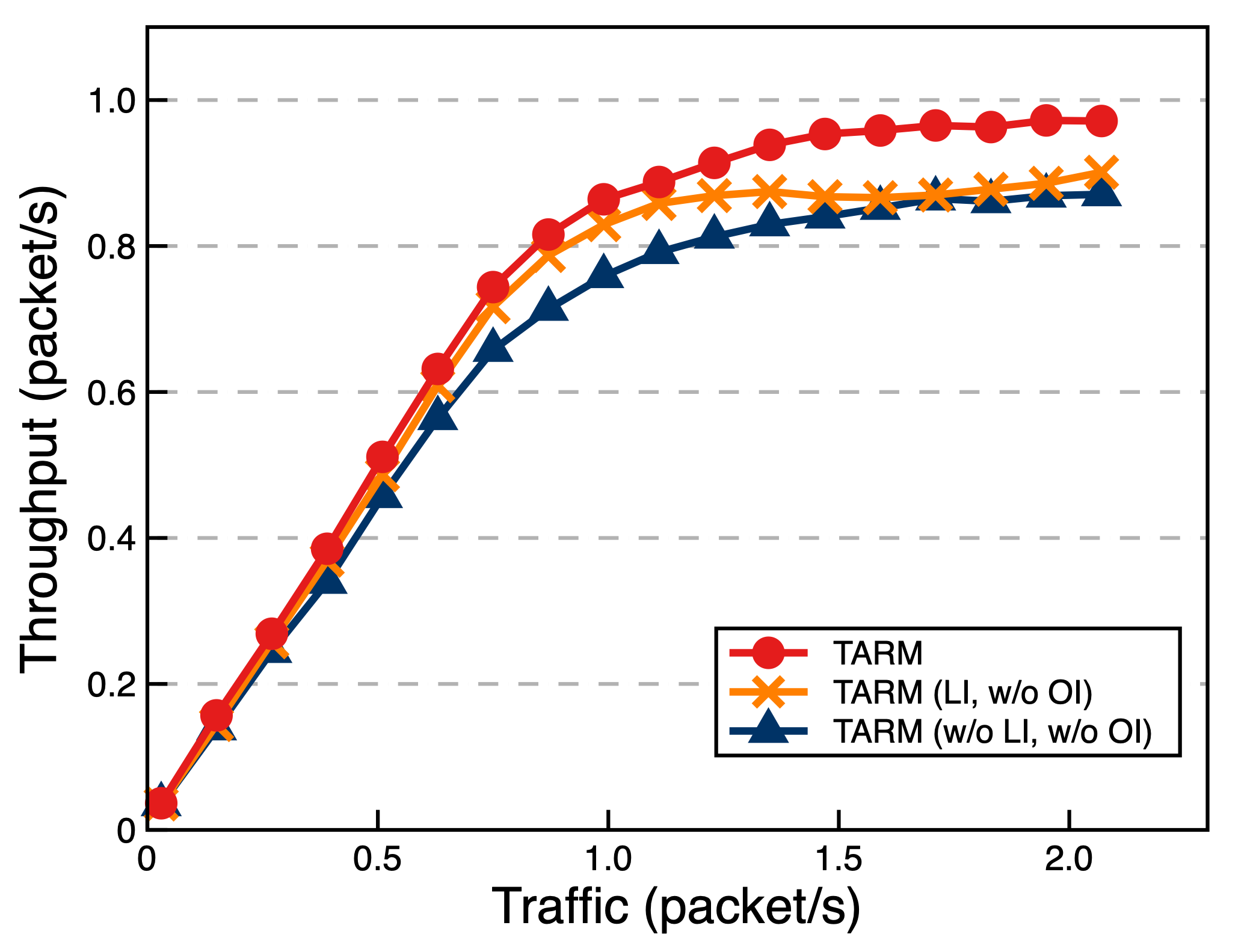}
\label{fig:TARM_comp_throughput}
\end{minipage}
}%
\subfigure[End-to-end delay]{
\begin{minipage}[t]{0.3\linewidth}
\centering
\includegraphics[width=2.2in]{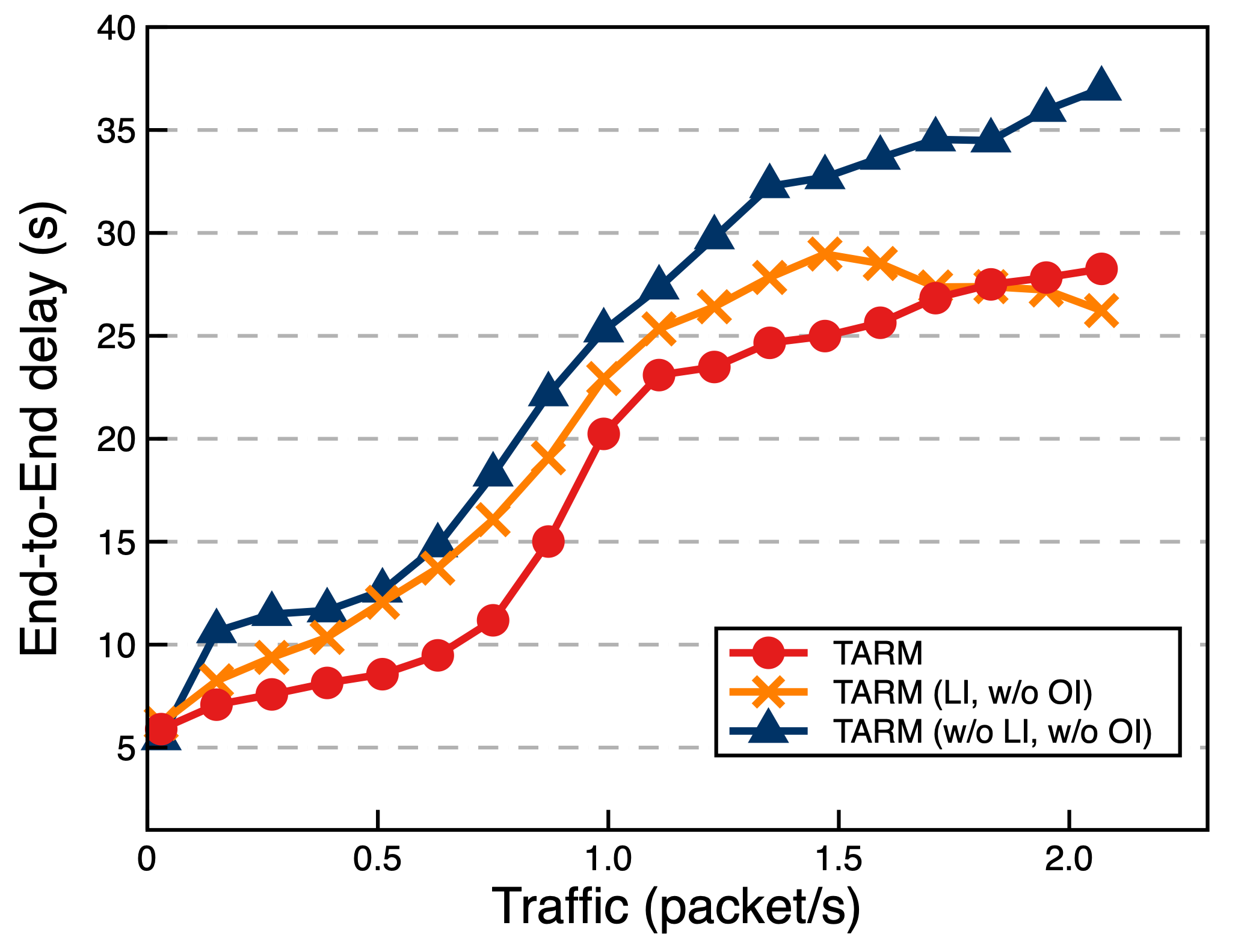}
\label{fig:TARM_comp_delay}
\end{minipage}
}%
\subfigure[Average energy consumption]{
\begin{minipage}[t]{0.3\linewidth}
\centering
\includegraphics[width=2.2in]{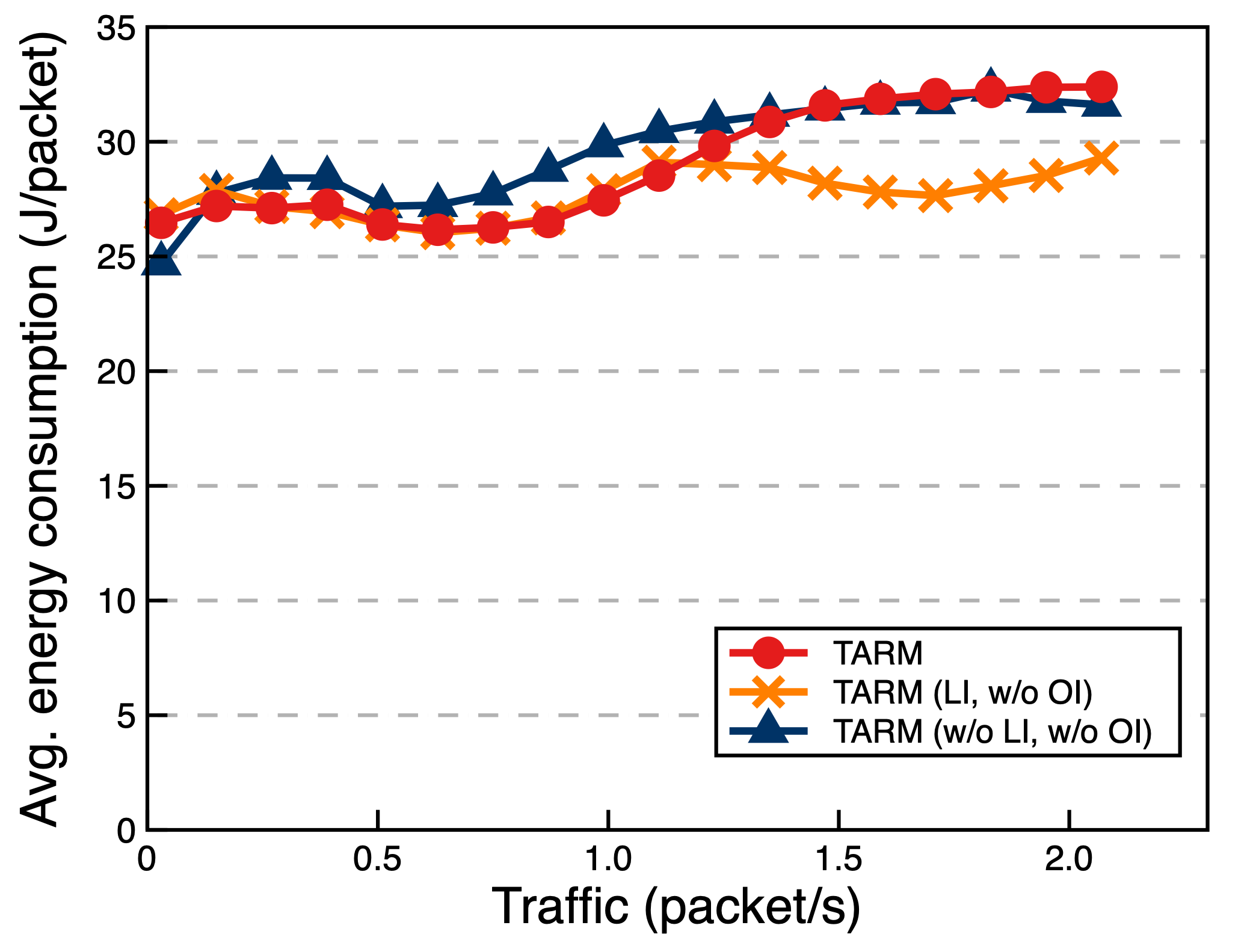}
\label{fig:TARM_comp_energy}
\end{minipage}
}%
\\
\subfigure[Delivery ratio]{
\begin{minipage}[t]{0.3\linewidth}
\centering
\includegraphics[width=2.2in]{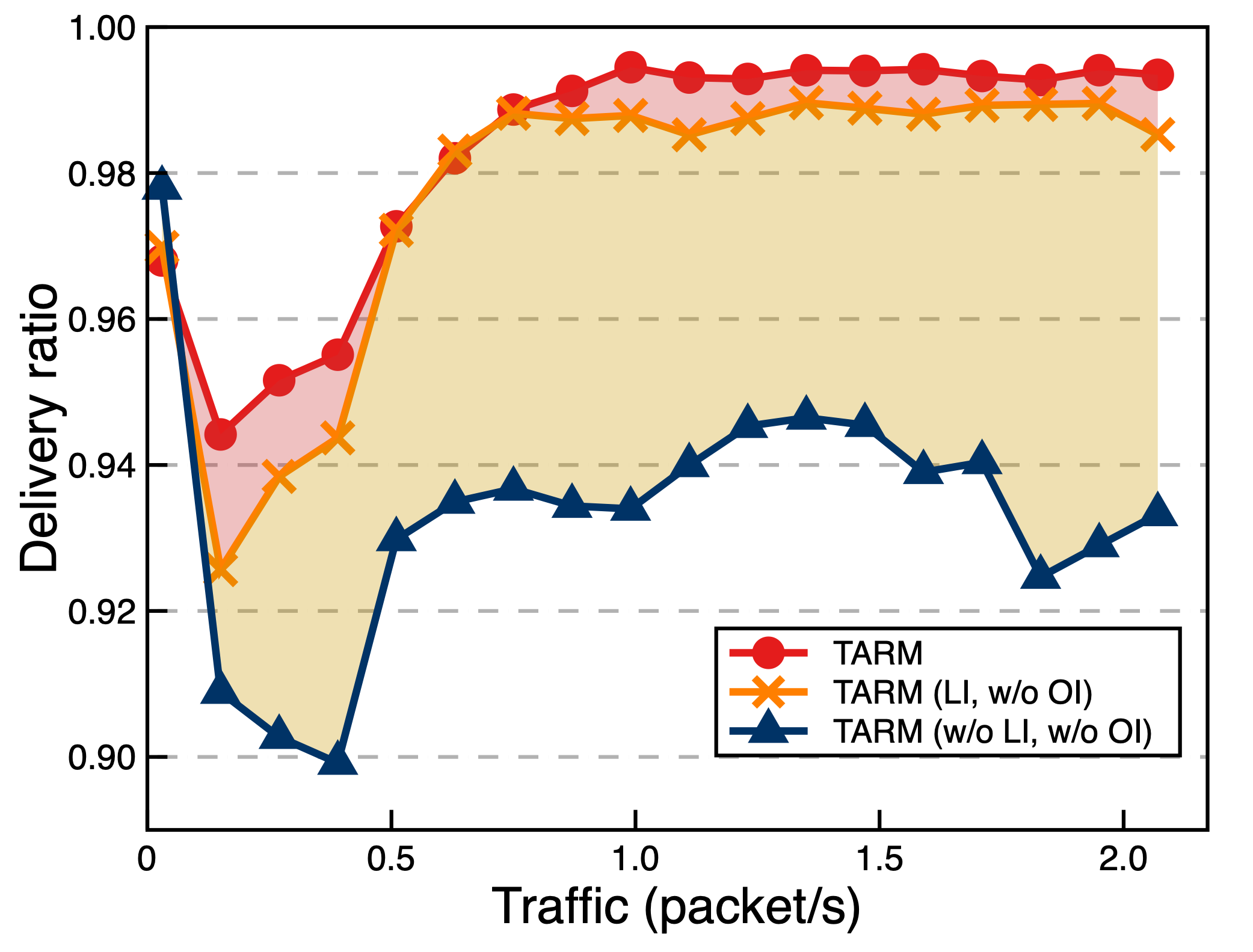}
\label{fig:TARM_comp_delivery}
\end{minipage}
}%
\subfigure[Channel utilization]{
\begin{minipage}[t]{0.3\linewidth}
\centering
\includegraphics[width=2.2in]{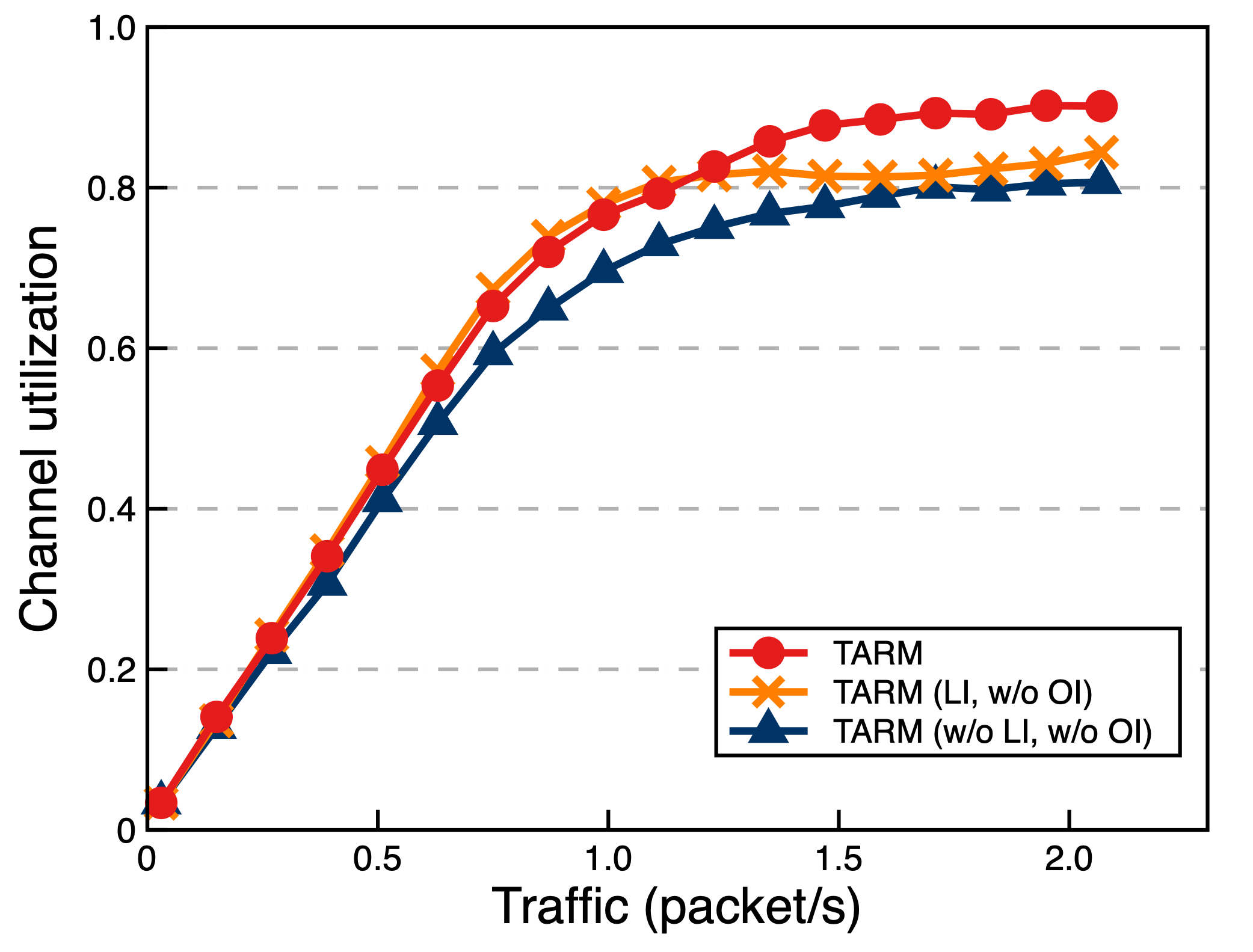}
\label{fig:TARM_comp_utilization}
\end{minipage}
}%
\centering
\caption{Impact of TARM components on network performance.}
\label{fig:TARM_comp}
\end{figure*}

From Fig. \ref{fig:TARM_packet_throughput}, several key observations can be made. First, as the packet length grows, TARM exhibits higher throughput and consistently outperforms other baseline methods. Second, in scenarios with longer packet lengths, leveraging the long propagation delay characteristics of underwater acoustic communication can significantly enhance channel utilization and optimize network throughput. In contrast, DR-DLMA, which also employs a reinforcement learning approach, experiences significant performance degradation when confronted with increasing packet lengths due to its inability to effectively exploit long propagation delays. Third, as the packet length increases, even when nodes are capable of transmitting packets at maximum rates, additional transmission time is required to process the augmented packet length, as illustrated in Figs. \ref{fig:TARM_packet_delay} and \ref{fig:delay_size}. The lack of a link scheduling mechanism results in a substantial number of communication failures stemming from conflicts, thus severely compromising overall network performance.

Figures \ref{fig:TARM_packet_delay} and \ref{fig:TARM_packet_energy} reveals noteworthy findings. Specifically, NF-TDMA and DR-DLMA exhibit reduced energy consumption per packet; however, this comes at the expense of increased transmission delays. On the other hand, TARM achieves a better trade-off between transmission delay and energy consumption. This is attributed to its adaptive adjustment of transmission parameters, which allows for optimized network throughput. Additionally, TARM employs traffic load-aware link scheduling decisions, effectively mitigating communication conflicts and achieving graceful degradation in packet delivery ratio as packet length increases.

\subsection{Impact of TARM Components on Network Performance}\label{sec:TARM_result_comp}

This section presents a validation of the essential design components employed by TARM through a selective omission of specific components, including 1) local load information $LI$, and 2) neighbor load information $OI$. The following variants of TARM are implemented for evaluation:
\begin{enumerate}[1)]
\item \textbf{TARM (LI, w/o OI):} Apart from the state and position information of the transmitter, nodes only use local load information as inputs to the decision-making model, without incorporating overhear information from neighboring nodes.
\item \textbf{TARM (w/o LI, w/o OI):} Nodes only employ their own transmission state and position information as inputs to the decision-making model, without including local load estimation or overhear information from neighboring nodes.
\end{enumerate}

\textcolor{blue}{For reference by neighboring nodes, the TARM node includes additional traffic information in each transmitted packet, making the packet length 200 bytes. However, both TARM (LI, w/o OI) and TARM (w/o LI, w/o OI) use smaller packet lengths of 190 bytes because they do not need to transmit local traffic estimates or overhear information to other nodes. Each simulation consists of 200 time slots, and the reported evaluation results are the average values derived from 100 simulation runs.}

\textcolor{blue}{In Fig. \ref{fig:TARM_comp}, it is evident that excluding any individual component from TARM leads to suboptimal network performance compared to the fine-tuned TARM strategy across different network traffic loads. This result has three important implications. Firstly, as network traffic gradually increases, all methods show a corresponding increase in network throughput. Importantly, all three methods demonstrate similar performance in throughput, with TARM slightly surpassing the other two variants. This observation underscores the effectiveness of the proposed MARL-based framework in optimizing network performance within partially observable systems. Additionally, it strengthens the efficacy of the TARM components in further enhancing network throughput under high-traffic conditions.}

\textcolor{blue}{Second, in high network traffic conditions, TARM components have a more significant impact on end-to-end delay. When network traffic remains below 0.63 pkt/s, there are minimal differences in end-to-end delay among the three methods. However, as traffic increases, the two TARM variants experience higher network delays.  This is because the decision-making process does not include information about the estimated traffic load from neighboring nodes. As a result, nodes with high traffic loads need assistance in obtaining adequate transmission opportunities to manage their traffic effectively, leading to increased delay for the TARM variants.}

Finally, integrating local load information into TARM significantly enhances network reliability, especially in high-traffic network scenarios. There are differences in the input lengths for the decision-making model among TARM and its simplified versions. TARM uses more information to make decisions, effectively reducing uncertainties caused by the mismatch between partial observations and global states. However, this may also add complexity to the decision-making model and increase communication overhead. Therefore, it is advisable to carefully choose the TARM components based on the specific performance objectives and constraints of the UWSNs.

\section{Conclusion}\label{sec:con}

This paper aims to provide efficient and reliable communication for underwater wireless sensor networks with limited energy and communication resources by optimizing communication link scheduling and adjusting transmission parameters such as transmit power and transmission rate. To achieve this, a traffic load-aware resource management strategy, denoted as TARM, has been proposed. TARM incorporates a traffic load-aware mechanism that uses information from neighboring nodes to mitigate discrepancies between partial observations and global states. Extensive simulations validate the effectiveness of TARM across various scenarios with different transmission demands and collision probabilities. The results demonstrate that TARM successfully supports efficient and reliable communications in these scenarios. It is important to note that this paper primarily focuses on a single-hop underwater communication scenario. Future work will explore the application of TARM in multi-hop UWSNs.

 \section*{Acknowledgments}
This work was supported by the National Key Research and Development Program (Grant No. 2021YFC2803000 and No. SQ2020YFB050001), the Joint Funds of the National Natural Science Foundation of China (Grant No. U22A2009).

\ifCLASSOPTIONcaptionsoff
  \newpage
\fi



%\bibliographystyle{IEEEtran}
\bibliographystyle{IEEEtran}
\bibliography{IEEEabrv,mybibfile}


\begin{IEEEbiography}[{\includegraphics[width=1in,height=1.25in,clip,keepaspectratio]{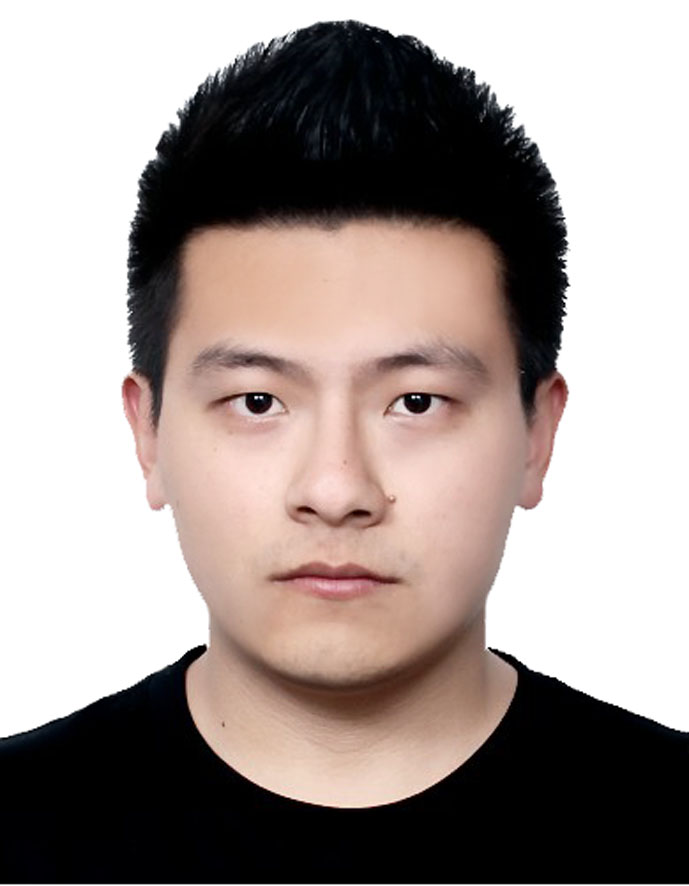}}]{Tong Zhang} received his B.S. degree (2014) from the College of Mathematics and Computer Science, Fuzhou University, Fuzhou, China, M.S. degree (2018) and PhD degree (2023) from the College of Computer Science and Technology, Jilin University, Changchun, China. He is currently a Post Doctoral Researcher with Beihang Ningbo Innovation Research Institute, Beihang University, Ningbo, China. His research interests include the resource management of UWSNs and deep reinforcement learning. He is a member of the IEEE Computer Society.
\end{IEEEbiography}
\begin{IEEEbiography}[{\includegraphics[width=1in,height=1.25in,clip,keepaspectratio]{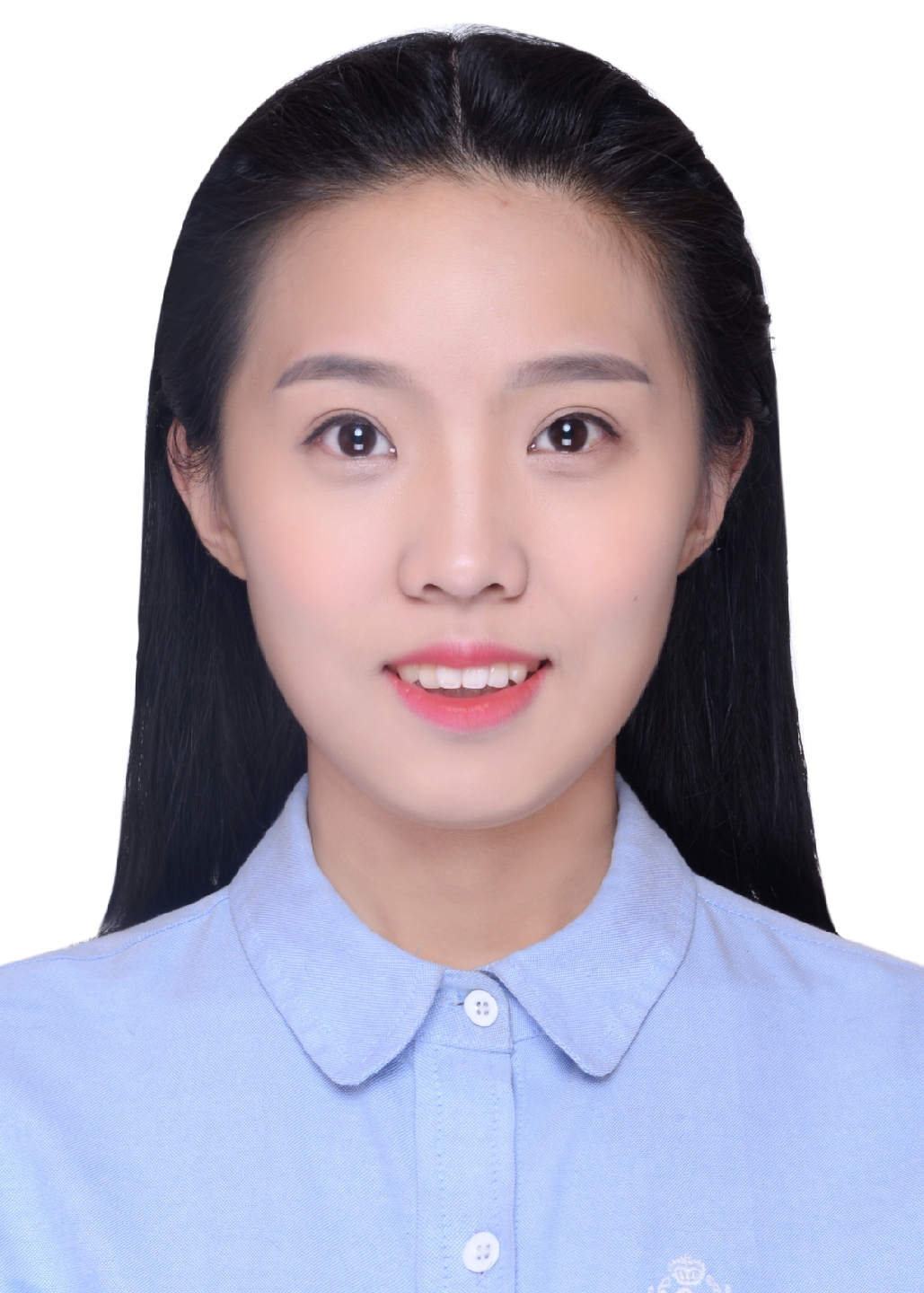}}]{Yu Gou} received her B.S. degree (2015), M.S. degree (2018), and PhD degree (2023) from the College of Computer Science and Technology, Jilin University, Changchun, China. She is currently a Post Doctoral Researcher with Beihang Ningbo Innovation Research Institute, Beihang University, Ningbo, China. Her research interests are include deep multi-agent reinforcement learning, underwater network performance optimization, and robust underwater networks. He is a member of the IEEE Computer Society.
\end{IEEEbiography}
\begin{IEEEbiography}[{\includegraphics[width=1in,height=1.25in,clip,keepaspectratio]{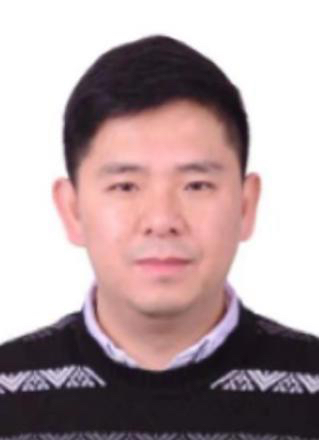}}]{Jun Liu} received the BEng degree (2002) in computer science from Wuhan University, China, the PhD degree (2013) in Computer Science and Engineering from University of Connecticut, USA. Currently, he is a professor of the School of Electronic and Information Engineering at Beihang University, China. His major research focuses on underwater networking, synchronization, and localization. He is a member of the IEEE Computer Society.
\end{IEEEbiography}
\begin{IEEEbiography}[{\includegraphics[width=1in,height=1.25in,clip,keepaspectratio]{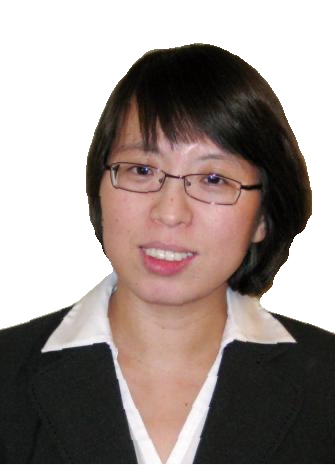}}]{Jun-Hong Cui} received the BS degree (1995) in computer science from Jilin University, China, the MS degree (1998) in computer engineering from Chinese Academy of Sciences, and the PhD degree (2003) in computer science from UCLA. Currently, she is on the faculty of the College of Computer Science and Technology, Jilin University, China. Recently, her research mainly focuses on exploiting the spatial properties in the modeling of network topology, algorithm and protocol design in underwater sensor networks.
\end{IEEEbiography}
%\newpage


%\vfill


\end{document}